\documentclass[12pt,eqsecnum]{article}
\usepackage[dvips]{graphicx}
\usepackage{amssymb}
\usepackage{amsmath}
\usepackage{epstopdf}
\makeatletter

\@addtoreset{equation}{section}
\makeatletter

\newcommand{\ma}[1]{\mbox{$\mathcal{#1}$}}
\newcommand{\qed}{\hbox{\rule[-2pt]{6pt}{6pt}}}
\newcommand{\D}{{\rm d}}

\newtheorem{Prop}{Proposition}

\newtheorem{lm}{Lemma}
\newtheorem{dn}{Definition}
\newtheorem{Coro}{Corollary}

\newcommand{\dalm}{\kern1pt\vbox{\hrule height 0.9pt\hbox{\vrule width
0.9pt\hskip 2.5pt\vbox{\vskip 5.5pt}\hskip 3pt\vrule width 0.3pt}\hrule height
0.3pt}\kern1pt}

\def\b2hat{ {\hat b}_2 }

\begin{document}

\begin{titlepage}
\vfill
\begin{flushright}
\today
\end{flushright}

\vfill
\begin{center}
\baselineskip=16pt
{\Large\bf 
Lovelock black holes with maximally symmetric horizons
}
\vskip 0.5cm
{\large {\sl }}
\vskip 10.mm
{\bf Hideki Maeda, Steven Willison and Sourya Ray} \\

\vskip 1cm
{
	Centro de Estudios Cient\'{\i}ficos (CECS), Casilla 1469, Valdivia, Chile \\
	\texttt{hideki@cecs.cl, willison@cecs.cl, ray@cecs.cl}

     }
\vspace{6pt}
\end{center}
\vskip 0.2in
\par
\begin{center}
{\bf Abstract}
 \end{center}
\begin{quote}
We investigate some properties of $n(\ge 4)$-dimensional spacetimes having symmetries corresponding to the isometries of an $(n-2)$-dimensional maximally symmetric space in Lovelock gravity under the null or dominant energy condition.
The well-posedness of the generalized Misner-Sharp quasi-local mass proposed in the past study is shown.
Using this quasi-local mass, we clarify the basic properties of the dynamical black holes defined by a future outer trapping horizon under certain assumptions on the Lovelock coupling constants.
The $C^2$ vacuum solutions are classified into four types: (i) Schwarzschild-Tangherlini-type solution; (ii) Nariai-type solution; (iii) special degenerate vacuum solution; (iv) exceptional vacuum solution.
The conditions for the realization of the last two solutions are clarified.
The Schwarzschild-Tangherlini-type solution is studied in detail.
We prove the first law of black-hole thermodynamics and present the expressions for the heat capacity and the free energy.
  \vfill
\vskip 2.mm
\end{quote}
\end{titlepage}




\tableofcontents

\section{Introduction}
\label{sec1}
Einstein's general theory of relativity is the most successful theory of
gravity which describes the movement of the planets in the solar system as
well as the dynamics of the expanding universe.
On the other hand, higher-dimensional gravity has been a prevalent
subject for the last decade.
This is strongly motivated by Superstring/M-theory which predicts the
existence of extra dimensions in our universe.
On the more Relativity side, this prevalence was triggered by the
remarkable discovery of the black-ring solution in five dimensions~\cite{er2002} which explicitly shows the richer structure of the higher-dimensional spacetime.
Undoubtedly, the most exciting prediction of general relativity is the
existence of spacetime regions from which nothing can escape, namely black
holes.
Black holes carry a lot of information about the gravity theory in
question, and the study of such objects is expected to yield insights into
the fundamental properties of gravity.

The Lovelock Lagrangian is the most natural generalization of the
Einstein-Hilbert Lagrangian for general relativity in arbitrary
dimensions without torsion such that the resulting field equations are
of second-order~\cite{lovelock}.
It consists of a sum of dimensionally extended Euler densities and
reduces to the Einstein-Hilbert Lagrangian with a cosmological constant
in four dimensions.
The second-order field equations in Lovelock gravity give rise to the
ghost-free
nature of the theory.
In string theory, the Einstein-Hilbert Lagrangian, which is the Lovelock
term linear in the curvature tensor, is realized as the lowest order term
in the Regge slope expansion of strings.
However, the forms of the higher-curvature terms appearing as the next-order stringy corrections depend on the type of string theories. (See~\cite{bentobertolami1996}, for example.)
Among them, the second-order Lovelock term, called the Gauss-Bonnet
term, appears in heterotic string theory~\cite{Gross}.
Apart from anything else, Lovelock gravity is of potential importance to
give insights how special four-dimensional spacetime is.

In dimensions greater than four, the Lovelock Lagrangian contains many coupling constants, one for each term
of particular order in the curvature.
This in turn implies that there are in general more than one maximally
symmetric vacuum solution such as Minkowski, de~Sitter (dS), or
anti-de~Sitter (AdS).
Interestingly, when we tune the coupling constants to admit a unique
vacuum, the Lagrangian of maximal order in the curvature has an additional
symmetry in odd dimensions and
becomes a gauge theory for the Poincar{\'e}, dS, or AdS group, where the
last two are the smallest nontrivial choices of such required groups
containing the translation symmetry group on a pseudosphere.
This resulting theory is called Chern-Simons gravity and has been
studied with keen interest due to aesthetic reasons.
(See~\cite{zanelli2005} for a review.)

However, for generic coupling constants, the higher curvature terms in
the field equations in Lovelock gravity make the analyses cumbersome.
In order to reduce the complexity, until now the research on Lovelock gravity has
been focused on spacetimes with high degrees of symmetry.
Especially, $n(\ge 4)$-dimensional spacetimes having symmetries
corresponding to the isometries of an $(n-2)$-dimensional maximally
symmetric space have been intensively investigated.
There is a Schwarzschild-Tangherlini-type vacuum black-hole solution and
the corresponding Jebsen-Birkhoff theorem was established for generic
coupling constants~\cite{zegers2005}.
Based on this solution, the thermodynamical aspects of static black
holes have been investigated~\cite{whitt1988,lovelock-thermo,cai2004}.
(See~\cite{lovelockreview} for a review of Lovelock black holes.)
Thermodynamical properties of more general concepts of horizons in Lovelock gravity were also studied~\cite{ck2005,ac2007,cc2007,cchk2008,Padmanabhan:2002jr,paddy2009}.
Very recently, it was shown that, unlike the four-dimensional
Schwarzschild black hole, sufficiently small higher-dimensional vacuum Lovelock
black holes are dynamically unstable in asymptotically flat spacetime
against tensor-type and scalar-type gravitational perturbations in even
and odd dimensions, respectively~\cite{ts2009}.
(See~\cite{takahashi2011} for the charged case.)
This remarkable result could suggest that the four-dimensional spacetime
is special.

At present, there are many important open problems in less symmetric
spacetimes in Lovelock gravity.
For example, the counterpart of the Myers-Perry rotating vacuum
solution is yet to be found~\cite{mp}.
(A five-dimensional rotating vacuum solution was found in the case of
Chern-Simons gravity~\cite{rotateCS}.)
Strong results in symmetric spacetimes will provide a firm basis for
the study of less symmetric spacetimes.
The present paper is written for this purpose.
In fact, the dynamical aspects of Lovelock black holes have not been
fully investigated so far.
In the quadratic case, namely in Einstein-Gauss-Bonnet gravity, it was
shown in the symmetric spacetime that the results in general relativity can be generalized in a
unified manner by introducing a well-defined quasi-local mass~\cite{maeda2006b,mn2008,nm2008,maeda2008,maeda2010}.
This allows to classify the solutions depending on whether they have
a general relativistic limit or not, and it was shown that the
solution which has a general relativistic limit has similar properties as
in general relativity, while the other has pathological properties.
In the present paper, those results are extended to general Lovelock
gravity.
Using the quasi-local mass in Lovelock gravity proposed in~\cite{mn2008}
, we prove a number of propositions in a similar fashion
to the general relativistic case.
We also present a number of new results.

The rest of the present paper is constituted as follows.
In the following section, we give an introduction to Lovelock gravity and
our spacetime ansatz.
In Section~III, we study the properties of the generalized Misner-Sharp
quasi-local mass proposed in~\cite{mn2008}.
Section~IV focuses on the study of dynamical black holes defined by a
future outer trapping horizon.
We also present several exact solutions with various kinds of matter fields as concrete examples of dynamical spacetimes with trapping horizons.
In section V, we classify all the vacuum solutions.
In section VI, we study the static vacuum black hole in more detail.
Concluding remarks and discussions including future prospects are
summarized in Section~VII.
In appendix A, we present expressions for the decomposition of geometric
tensors in our symmetric spacetime.
In appendix B, we give a brief comment on the case with a unique maximally symmetric vacuum.
In appendix C, we discuss about the Schwarzschild-Tangherlini-type solution with electric charge.

Our basic notation follows \cite{wald}.
The conventions for curvature tensors are
$[\nabla _\rho ,\nabla_\sigma]V^\mu ={R^\mu }_{\nu\rho\sigma}V^\nu$
and $R_{\mu \nu }={R^\rho }_{\mu \rho \nu }$.
The Minkowski metric is taken to be the mostly plus sign, and
Roman indices run over all spacetime indices.
We adopt the units in which only the $n$-dimensional gravitational
constant $G_n$ is retained.

~\\ {\em Note added:}

At the final stage of the present work we were informed that X.O.~Camanho and J.D.~Edelstein were working on a similar subject~\cite{ce2011}.

\section{Preliminaries}
\label{sec2}
\subsection{Lovelock gravity}
We begin with an introduction to Lovelock gravity.
The Lovelock action in $n (\geq 4)$-dimensional spacetime is given by
\begin{align}
\label{action}
I=&\frac{1}{2\kappa_n^2}\int \D ^nx\sqrt{-g}\sum_{p=0}^{[n/2]}\alpha_{(p)}{\ma L}_{(p)}+I_{\rm matter},\\
{\ma L}_{(p)}:=&\frac{1}{2^p}\delta^{\mu_1\cdots \mu_p\nu_1\cdots \nu_p}_{\rho_1\cdots \rho_p\sigma_1\cdots \sigma_p}R_{\mu_1\nu_1}^{\phantom{\mu_1}\phantom{\nu_1}\rho_1\sigma_1}\cdots R_{\mu_p\nu_p}^{\phantom{\mu_p}\phantom{\nu_p}\rho_p\sigma_p},
\end{align}
where $\kappa_n := \sqrt{8\pi G_n}$.
We assume $\kappa_n^2>0$ without any loss of generality and make no assumptions about the signs of $\alpha_{(p)}$ unless otherwise mentioned.
The $\delta$ symbol denotes a totally anti-symmetric product of Kronecker deltas, normalized to take values $0$ and $\pm 1$~\cite{lovelock,km2006}, defined by
\begin{align}
\delta^{\mu_1\cdots \mu_p}_{\rho_1\cdots \rho_p}:=&p!\delta^{\mu_1}_{[\rho_1}\cdots \delta^{\mu_p}_{\rho_p]}.
\end{align}
$\alpha_{(p)}$ is a coupling constant with dimension $({\rm length})^{2(p-1)}$.
$\alpha_{(0)}$ is related to the cosmological constant $\Lambda$ by $\alpha_{(0)}=-2\Lambda$.
The general Lovelock Lagrangian density is given by an arbitrary linear combination of dimensionally continued Euler densities. In even dimension $n$, the variation of the $n$-dimensional Euler density is a total derivative and does not contribute to the field equations and in any dimension $n$, the $(n+1)$-dimensional Euler density vanishes identically. Therefore the sum truncates at a finite order.
The first four Lovelock Lagrangians are explicitly shown as
\begin{align}
{\ma L}_{(0)}:=& 1,\\
{\ma L}_{(1)}:=& R,\\
{\ma L}_{(2)}:=& R^2-4R_{\mu\nu}R^{\mu\nu}+R_{\mu\nu\rho\sigma}R^{\mu\nu\rho\sigma},\\
{\ma L}_{(3)}:=& R^3   -12RR_{\mu \nu } R^{\mu \nu } + 16R_{\mu \nu }R^{\mu }_{\phantom{\mu } \rho }R^{\nu \rho }+ 24 R_{\mu \nu }R_{\rho \sigma }R^{\mu \rho \nu \sigma }+ 3RR_{\mu \nu \rho \sigma } R^{\mu \nu \rho \sigma } \nonumber \\
&-24R_{\mu \nu }R^\mu _{\phantom{\mu } \rho \sigma \kappa } R^{\nu \rho \sigma \kappa  }+ 4 R_{\mu \nu \rho \sigma }R^{\mu \nu \eta \zeta } R^{\rho \sigma }_{\phantom{\rho \sigma } \eta \zeta }-8R_{\mu \rho \nu \sigma } R^{\mu  \phantom{\eta } \nu }_{\phantom{\mu } \eta 
\phantom{\nu } \zeta } R^{\rho  \eta  \sigma  \zeta }.
\end{align}
In even dimensions, the contribution to the action of the $n/2$-th order Lagrangian becomes a topological invariant and does not contribute to the field equations.

The gravitational equation following from this action is given by 
\begin{align} 
{\ma G}_{\mu\nu}=\kappa_n^2 {T}_{\mu\nu}, \label{beqL}
\end{align} 
where ${T}^\mu_{~~\nu}$ is the energy-momentum tensor for matter fields obtained from $I_{\rm matter}$ and
\begin{align} 
{\ma G}_{\mu\nu} :=& \sum_{p=0}^{[n/2]}\alpha_{{(p)}}{G}^{(p)}_{\mu\nu}, \label{generalG}\\
{G}^{\mu(p)}_{~~\nu}:=& -\frac{1}{2^{p+1}}\delta^{\mu\eta_1\cdots \eta_p\zeta_1\cdots \zeta_p}_{\nu\rho_1\cdots \rho_p\sigma_1\cdots \sigma_p}R_{\eta_1\zeta_1}^{\phantom{\eta_1}\phantom{\zeta_1}\rho_1\sigma_1}\cdots R_{\eta_p\zeta_p}^{\phantom{\eta_p}\phantom{\zeta_p}\rho_p\sigma_p}.
\end{align} 
The tensor ${G}^{(p)}_{\mu\nu}$ is given from ${\ma L}_{(p)}$.
${G}^{(p)}_{\mu\nu}\equiv 0$ is satisfied for $p\ge [(n+1)/2]$.
There is an identity between the Lovelock Lagrangian and the Lovelock tensor:
\begin{align}
G^{\mu(p)}_{~~\mu} \equiv  \frac{2p-n}{2}{\cal L}_{(p)}.\label{id-L}
\end{align}
The field equations (\ref{beqL}) contain up to the second derivatives 
of the metric.
The first four Lovelock tensors are explicitly shown as
\begin{align}
{G}^{(0)}_{\mu\nu} =& -\frac12 g_{\mu\nu},\\
{G}^{(1)}_{\mu\nu}=&R_{\mu\nu}-\frac12 Rg_{\mu\nu},\\
{G}^{(2)}_{\mu\nu}=& 2\biggl(RR_{\mu\nu}-2R_{\mu\rho}R^\rho_{\phantom{\rho}\nu}-2R^{\rho\sigma}R_{\mu\rho\nu\sigma}+R_{\mu}^{\phantom{\mu}\rho\sigma\gamma}R_{\nu\rho\sigma\gamma}\biggl)-\frac12 g_{\mu\nu}{\ma L}_{(2)},\\
{G}^{(3)}_{\mu\nu}=&3\biggl(R^2 R_{\mu \nu } - 4 R R_{\rho  \mu } R^\rho _{\phantom{\rho }\nu } - 4R^{\rho \sigma }R_{\rho \sigma }R_{\mu \nu }+8 R^{\rho \sigma }R_{\rho  \mu }R_{\sigma  \nu } - 4 R R^{\rho \sigma }R_{\rho  \mu  \sigma  \nu }   \nonumber \\
&  +8 R^{\rho \kappa }R^\sigma _{\phantom{\sigma }\kappa }R_{\rho  \mu  \sigma  \nu } -16 R^{\rho \sigma }R^\kappa _{\phantom{\kappa } (\mu }R_{ |\kappa \sigma \rho |  \nu ) }+ 2 R R^{\rho \sigma \kappa }_{\phantom{\rho \sigma \kappa }\mu  }R_{\rho \sigma \kappa  \nu } +R_{\mu \nu }R^{\rho \sigma \kappa \eta } R_{\rho \sigma \kappa \eta  }   \nonumber \\
&- 8 R^\rho _{\phantom{\rho }(\mu }R^{\sigma \kappa \eta }_{\phantom{\sigma \kappa \eta } |\rho | }R_{|\sigma \kappa \eta | \nu ) }- 4 R^{\rho \sigma }R^{\kappa \eta }_{\phantom{\kappa \eta } \rho \mu }R_{\kappa \eta  \sigma  \nu }+8R_{\rho \sigma }R^{\rho  \kappa  \sigma  \eta }R_{\kappa  \mu  \eta  \nu } - 8 R_{\rho \sigma }R^{\rho  \kappa \eta }_{\phantom{\rho  \kappa \eta }\mu }R^\sigma _{\phantom{\sigma } \kappa \eta  \nu } \nonumber \\
&+4 R^{\rho \sigma  \kappa \eta }R_{\rho \sigma  \zeta  \mu  }R_{\kappa \eta  \phantom{\zeta } \nu }^{\phantom{\kappa \eta }\zeta } - 8 R^{\rho  \kappa  \sigma  \eta }R^\zeta _{\phantom{\zeta }\rho \sigma  \mu }R_{\zeta  \kappa \eta  \nu }- 4R^{\rho \sigma \kappa }_{\phantom{\rho \sigma \kappa } \eta  } R_{\rho \sigma \kappa  \zeta }R^{\eta  \phantom{\mu }\zeta }_{\phantom{\eta } \mu  \phantom{\zeta } \nu }\biggl)-  \frac12 g_{\mu \nu }{\cal L}_{(3)}.
\end{align}

\subsection{Lovelock tensors in the symmetric spacetime}
Consider an $n(\ge 4)$-dimensional spacetime $({\ma M}^n, g_{\mu \nu })$ to be a warped product of an 
$(n-2)$-dimensional maximally symmetric space $(K^{n-2}, \gamma _{ij})$ with sectional curvature $k = \pm 1, 0$
and a two-dimensional orbit spacetime $(M^2, g_{ab})$ under the isometries of $(K^{n-2}, \gamma _{ij})$. 
$\gamma_{ij}$ is the unit metric on $(K^{n-2}, \gamma _{ij})$.
The curvature of the $(n-2)$-dimensional maximally symmetric space is given by 
\begin{eqnarray}
{}^{(n-2)}{R}{}_{ijkl}=k(\gamma_{ik}\gamma_{jl}-\gamma_{il}\gamma_{jk}),
\end{eqnarray}
where the superscript $(n-2)$ means the geometrical quantity on $(K^{n-2}, \gamma _{ij})$.
We also assume that $(K^{n-2}, \gamma _{ij})$ is compact in order to have a finite value for certain physical quantities.

For such a spacetime, without loss of generality the line element may be given by
\begin{align}
g_{\mu \nu }\D x^\mu \D x^\nu =g_{ab}(y)\D y^a\D y^b +r^2(y) \gamma _{ij}(z)
\D z^i\D z^j ,
\label{eq:ansatz}
\end{align} 
where $a,b = 0, 1;~i,j = 2, ..., n-1$ and $r$ is a scalar on $(M^2, g_{ab})$. 
Using the expressions of the decomposed geometric tensors given in Appendix~A, the components of the $p$-th order Lovelock tensor tangent to $(M^2, g_{ab})$ and the $p$-th order Lovelock Lagrangian ${\cal L}_{(p)}$ are respectively given by
\begin{align}
G^{(p)}_{ab} = & -\frac{p(n-2)!}{(n-2p-1)!}\ \frac{D_aD_br-(D^2r)g_{ab}}{r}\left(  \frac{k-(Dr)^2 }{r^2} \right)^{p-1} \nonumber \\
&- \frac{(n-2)!}{2(n-2p-2)!}g_{ab}\left(  \frac{k- (Dr)^2}{r^2} \right)^{p},\label{Gab} \\
 {\cal L}_{(p)} =& \frac{(n-2)!}{(n-2p)!} \Psi_{(p)},\label{Lag}
\end{align}
where
\begin{align}
 \Psi_{(p)} :=&(n-2p)(n-2p-1) \left(  \frac{k- (Dr)^2}{r^2} \right)^{p}-2(n-2p) p  \frac{(D^2 r)}{r} \left(  \frac{k-(Dr)^2 }{r^2} \right)^{p-1} \nonumber\\
 &+ 2 p(p-1)\frac{(D^2 r)^2 - (D^a D_b r)(D^bD_a r)}{r^2}\left(  \frac{k-(Dr)^2 }{r^2} \right)^{p-2}+  p\overset{(2)}{R} \left(  \frac{k-(Dr)^2 }{r^2} \right)^{p-1}.
\end{align}
Here $D_a$ is a metric compatible linear connection on $(M^2, g_{ab})$, $(Dr)^2:=g^{ab}(D_ar)(D_br)$, and $D^2r:=D^aD_ar$.
The contraction was taken over the two-dimensional orbit space and ${}^{(2)}{R}$ is the Ricci scalar on $(M^2, g_{ab})$.

The $(i,j)$ component of the $p$-th order Lovelock tensor $G^{(p)}_{ij}$ is given by 
\begin{align}
G^{i(p)}_{~~j}=&-\frac{(n-3)!}{2(n-2p-1)!}\Phi_{(p)}\delta^i_{~~j},\label{Gij}
\end{align}
where 
\begin{align}
\label{nasty1}
\Phi_{(p)} :=& (n-2p-1)(n-2p-2) \left(  \frac{k- (Dr)^2}{r^2} \right)^{p} \nonumber \\
&-p\biggl\{2(n-2p-1) \frac{(D^2 r)}{r}-\overset{(2)}{R}{} \biggl\}\left(  \frac{k-(Dr)^2 }{r^2} \right)^{p-1} \nonumber\\
&+ 2  p(p-1) \frac{(D^2 r)^2 - (D^a D_b r)(D^bD_a r)}{r^2}\left(  \frac{k-(Dr)^2 }{r^2} \right)^{p-2}.
\end{align}
With Eqs.~(\ref{Gab}), (\ref{Lag}), and (\ref{Gij}), we can confirm the identity~(\ref{id-L}).
The $(i,j)$ component of the left-hand side of the field equation (\ref{beqL}) is
\begin{align}
{\cal G}^{i}_{~~j}=&-\sum_{p=0}^{[n/2]}\frac{p{\tilde\alpha}_{(p)}}{2}\left(  \frac{k-(Dr)^2 }{r^2} \right)^{p-2}{\tilde\Phi}_{(p)}\delta^i_{~~j},\label{besij}
\end{align}
where 
\begin{align}
{\tilde \alpha}_{(p)}:=&\frac{(n-3)!\alpha_{(p)}}{(n-1-2p)!}, \label{alphatil}\\
{\tilde\Phi}_{(p)} :=& \frac{(n-2p-1)(n-2p-2)}{p} \left(  \frac{k- (Dr)^2}{r^2} \right)^{2}-2(n-2p-1) \frac{(D^2 r)}{r} \left(  \frac{k-(Dr)^2 }{r^2} \right) \nonumber\\
&+ 2 (p-1) \frac{(D^2 r)^2 - (D^a D_b r)(D^bD_a r)}{r^2}+ \overset{(2)}{R}{} \left(  \frac{k-(Dr)^2 }{r^2} \right).\label{nasty2}
\end{align}
We note that ${\tilde \alpha}_{(n/2)} \equiv 0$ for even $n$ by definition.

Hence, the most general energy-momentum tensor $T_{\mu\nu}$ compatible with this spacetime symmetry governed by Lovelock equations is given by
\begin{align}
T_{\mu\nu}\D x^\mu \D x^\nu =T_{ab}(y)\D y^a\D y^b+p(y)r^2 \gamma_{ij}\D z^i\D z^j,
\end{align}  
where $T_{ab}(y)$ and $p(y)$ are a symmetric two-tensor and a scalar on $(M^2, g_{ab})$, respectively.

We note that the results in the present paper are valid for $k=0$ and $p=0$ when we set $k^p=1$.

\section{Generalized Misner-Sharp quasi-local mass}
\label{sec3}
In this section, we show that the generalized Misner-Sharp quasi-local mass proposed in~\cite{mn2008} is certainly a natural counterpart of the Misner-Sharp mass in general relativity~\cite{ms1964}.
The generalized Misner-Sharp mass in Lovelock gravity was defined as
\begin{align}
\label{qlm}
m_{\rm L} :=& \frac{(n-2)V_{n-2}^{(k)}}{2\kappa_n^2}\sum_{p=0}^{[n/2]}{\tilde \alpha}_{(p)}r^{n-1-2p}[k-(Dr)^2]^p,
\end{align}  
where $V_{n-2}^{(k)}$ denotes the area of $(K^{n-2}, \gamma _{ij})$~\cite{mn2008}.
In the four-dimensional spherically symmetric case 
without a cosmological constant, 
$m_{\rm L}$ reduces to the Misner-Sharp quasi-local mass~\cite{ms1964}.
We assume the reality of $m_{\rm L}$ throughout the paper.

\subsection{Unified first law}
First, it is shown that the components of the Lovelock equation on $(M^2, g_{ab})$ can be written in the form of the so-called unified first law using $m_{\rm L}$.
This unified first law was first shown in~\cite{tw2011} adopting a coordinate system on $(M^2, g_{ab})$.
Here we present a tensorial proof.
\begin{Prop}
\label{th:1stlaw}
({\it Unified first law in Lovelock gravity.}) 
The following unified first law holds:
\begin{align}
\D  m_{\rm L} =&{\cal A}\psi_a\D x^a  +P\D V, \label{1stlaw1}
\end{align}
where
\begin{align}
\psi_a :=&{T_a}^bD_b r +PD_a r,\quad P:=-\frac 12 {T^a}_a, \\
V:=&\frac{V_{n-2}^{(k)}r^{n-1}}{n-1},\quad {\cal A}:=V_{n-2}^{(k)}r^{n-2}. \label{area}
\end{align}
\end{Prop}
\noindent
{\it Proof}.
The component of the field equation (\ref{beqL}) on $(M^2, g_{ab})$ is 
\begin{align}
\kappa_n^2T_{ab} = & -\sum_{p=0}^{[n/2]}p(n-2){\tilde\alpha}_{(p)}\ \frac{D_aD_br-(D^2r)g_{ab}}{r}\left(  \frac{k-(Dr)^2 }{r^2} \right)^{p-1} \nonumber \\
&- \sum_{p=0}^{[n/2]}{\tilde \alpha}_{(p)}\  \frac{(n-2p-1)(n-2)}{2}g_{ab}\left(  \frac{k- (Dr)^2}{r^2} \right)^{p}. \label{comp}
\end{align}
The following expressions are useful:
\begin{align}
G^{(p)}_{ab}(D^b r)- G^{(p)b}_{b}(D_ar) = &-\frac{p(n-2)!}{(n-2p-1)!}\ \frac{(D_aD_br)(D^b r)}{r}\left(  \frac{k-(Dr)^2 }{r^2} \right)^{p-1} \nonumber\\
&+ \frac{(n-2)!}{2(n-2p-2)!}\left(  \frac{k- (Dr)^2}{r^2} \right)^{p}(D_a r),\label{sub1} \\  
G^{(p)}_{ab}- \frac{1}{2} g_{ab}G^{(p)d}_{d} = & -\frac{p(n-2)!}{(n-2p-1)!}\ r^{-1}\biggl(D_aD_br-\frac12 (D^2r)g_{ab}\biggl)\left(  \frac{k-(Dr)^2 }{r^2} \right)^{p-1}.\label{beq-simp}
\end{align}
The derivative of $m_{\rm L}$ gives
\begin{align}
D_a m_{\rm L} =&\frac{V_{n-2}^{(k)}r^{n-2}}{\kappa_n^2}\sum_{p=0}^{[n/2]}\biggl(-\frac{p(n-2)!\alpha_{(p)}}{(n-1-2p)!}r^{1-2p}[k-(Dr)^2]^{p-1}(D_a D_br)(D^b r) \nonumber\\
&+\frac{(n-2)!\alpha_{(p)}}{2(n-2-2p)!}r^{-2p}[k-(Dr)^2]^p(D_a r)\biggl),\label{sub2}  \\
=&\frac{V_{n-2}^{(k)}r^{n-2}}{\kappa_n^2}\sum_{p=0}^{[n/2]}\alpha_{(p)}\biggl(G^{(p)}_{ab}(D^b r)- G^{(p)b}_{b}(D_ar) \biggl),\label{ufl-D}
\end{align}  
where we used Eq.~(\ref{sub1}).
Using the fied equation (\ref{beqL}), we obtain the unified first law (\ref{1stlaw1}) from Eq.~(\ref{ufl-D}).

\qed

\bigskip

The volume $V$ of $(K^{n-2}, \gamma _{ij})$ and the surface area ${\cal A}$ satisfy $D_a V={\cal A}D _ar$.
Equation (\ref{1stlaw1}) does not contain the Lovelock coupling constants explicitly and is of the same form as in the general relativistic case~\cite{hayward1998}.
The unified first law will be used to derive the dynamical entropy of a black hole in the subsequent section.

\subsection{Quasi-local mass and the Kodama vector}
In this subsection we show that the generalized Misner-Sharp mass is interpreted as a locally conserved quantity.
We define the locally conserved current vector $J^\mu$ as
\begin{align}
J^\mu :=& -\frac{1}{\kappa _n^2}{\cal G}^\mu_{~~\nu}K^\nu=-{T^\mu}_\nu K^\nu, \label{j} \\
K^\mu  :=&-\epsilon ^{\mu \nu }\nabla _\nu  r. \label{kodamavector}
\end{align}
Here $\epsilon_{\mu \nu}=\epsilon_{ab}(\D x^a)_{\mu}(\D x^b)_{\nu}$, and $\epsilon_{ab}$ is a volume element of $(M^2, g_{ab})$. 
$K^\mu$ is the Kodama vector~\cite{kodama1980} and satisfies
\begin{align}
K^\mu K_\mu=-(D r)^2.\label{ksquare}
\end{align}
Hence, it is timelike (spacelike) in the untrapped (trapped) region.
$K^\mu$ is divergence-free and generates a preferred time evolution vector field in the untrapped region.
(See also~\cite{av2010}.)
In a static spacetime, $K^\mu$ reduces to the hypersurface-orthogonal timelike Killing vector.
Also, $J^\mu$ is divergence-free because of the identity $\nabla_{\nu}{\cal G}^{\mu\nu}\equiv 0$.
Indeed, $m_{\rm L}$ is a quasi-local conserved quantity associated with a locally conserved current $J^\mu$, which is seen in the following expression of $J^\mu$:
\begin{align}
J^\mu =-\frac{1}{V^{(k)}_{n-2}}r^{-(n-2)}\epsilon^{\mu \nu}\nabla_\nu m_{\rm L}.  \label{j-m}
\end{align}
(Similar expressions to (\ref{j}) and (\ref{j-m}) were obtained also in the stationary case given in section 7 in~\cite{kastor2008}.) 
The above equation is obtained as follows.
Using the identity $\delta^a_c \delta^b_d - \delta^b_c \delta^a_d =-\epsilon^{ab}\epsilon_{cd}$, it follows from equation (\ref{ufl-D}) that 
\begin{align}
D_a m_{\rm L}=-V^{(k)}_{n-2}r^{n-2}\epsilon_{ab}\epsilon^{cd}T^{b}_{\phantom{b}c} D_d r.
\end{align}
Furthermore, using the identity $\epsilon_{ab}\epsilon^{bc} = \delta^c_a$
we obtain
\begin{align}
\epsilon^{ab}D_b m_{\rm L}=-V^{(k)}_{n-2}r^{n-2}\epsilon^{cd}T^{a}_{\phantom{a}c} D_d r.
\end{align}
Hence, we obtain
\begin{align}
J^a =\epsilon ^{cd}{T^a}_c D_d  r =-\frac{1}{V^{(k)}_{n-2}}r^{-(n-2)}\epsilon^{ab}D_bm_{\rm L},
\end{align}
which shows Eq.~(\ref{j-m}). From Eq.~(\ref{j-m}), we immediately obtain
\begin{align}
{\cal L}_J m_{\rm L}=&J^\mu \nabla_\mu m_{\rm L}=0,\label{kodamamass}
\end{align}
which implies that $m_{\rm L}$ is conserved along $J^\mu$.
As a consequence, the integral of $J^\mu $ over some spatial volume with boundary $\Sigma $ gives $m_{\rm L}$ as an associated charge as $m_{\rm L}= \int _\Sigma J^\mu \D \Sigma _\mu$, where $\D \Sigma _\mu$ is a directed surface element on $\Sigma $.
(See section III in~\cite{mn2008}.)

It can be shown for static spacetimes that $\psi^a$ is vanishing and the unified first law (\ref{1stlaw1}) takes a simpler form.
Using Eq.~(\ref{beq-simp}), we obtain
\begin{align}
\kappa_n^2 \psi_a = & (n-2)\frac{(D^2r)D_a r-2(D_aD_br)D^b r}{2r}\sum_{p=0}^{[n/2]}p{\tilde\alpha}_{(p)}\left(  \frac{k-(Dr)^2 }{r^2} \right)^{p-1}.\label{psi-2}
\end{align}
Using the identity $\epsilon_{ab}\epsilon^{cd}=-2{\delta^c}_{[a}{\delta^d}_{b]}$, we obtain
\begin{align}
K^bD_{(b}K_{a)} &=\frac12 (D^2 r)D _a r-(D^br)D_bD_a r.\label{derivk2}
\end{align}
Comparing Eqs.~(\ref{psi-2}) and (\ref{derivk2}), we conclude that $\psi^a=0$ in static spacetimes, in which $K^\mu$ is a Killing vector and hence $D_{(b}K_{a)}=0$.
This results surely implies that $\psi^a$ represents an energy flux.

\subsection{Global mass, monotonicity, positivity, and rigidity}
\label{subsec:qlm}
In this subsection, we show several important properties of the generalized Misner-Sharp mass.
First it is shown that the quasi-local mass $m_{\rm L}$ asymptotes to the Arnowitt-Deser-Misner (ADM) mass, calculated using the same geometrical formula as used in general relativity, in asymptotically flat spacetime.
(The proof is similar to Proposition 2 in~\cite{mn2008} since the higher-order terms converge to zero more rapidly at spatial infinity.
)
\begin{Prop}
\label{th:asymptotics}
({\it Asymptotic behavior in asymptotically flat spacetime.}) 
In an $n(\ge 4)$-dimensional asymptotically flat spacetime, 
$m_{\rm L}$ coincides with the higher-dimensional ADM mass at spatial infinity.
\end{Prop}

\bigskip

Hereafter, we adopt the double-null coordinates on $(M^2,g_{ab})$ as
\begin{align}
\D s^2 = -2e^{-f(u,v)}\D u\D v+r^2(u,v) \gamma_{ij}\D z^i\D z^j. \label{coords}
\end{align}  
Null vectors $(\partial /\partial u)$ and $(\partial /\partial v)$ are taken to be future-pointing. 
The area expansions along two independent future-directed null vectors $(\partial /\partial u)$ and $(\partial /\partial v)$ are defined as
\begin{align}
\theta_{+}&:=\frac{{\ma L}_v {\cal A}}{{\cal A}}=(n-2)r^{-1}\partial_{v}r,\\
\theta_{-}&:=\frac{{\ma L}_u {\cal A}}{{\cal A}}=(n-2)r^{-1}\partial_{u}r,
\end{align}  
where ${\cal A}$ is the area of the symmetric subspace defined by Eq.~(\ref{area}).
They are used in the following definition.
\begin{dn}
A {\it trapped (untrapped) surface} is a compact $(n-2)$-surface with $\theta_{+}\theta_{-}>(<)0$.
A {\it trapped (untrapped) region} is the union of all trapped (untrapped) surfaces.
A {\it marginal surface} is a compact $(n-2)$-surface with $\theta_{+}\theta_{-}=0$.
\end{dn}
In this section, we fix the orientation of the untrapped surface by
$\theta _+>0$ and $\theta_-<0$, i.e., $\partial/\partial u$ and $\partial/\partial v$ are ingoing and outgoing null vectors, respectively.
With this orientation, the Kodama vector
\begin{align}
K^\mu\frac{\partial}{\partial x^\mu}=e^{f}D_v r \frac{\partial}{\partial u}-e^{f}D_u r\frac{\partial}{\partial v} \label{kodama2}
\end{align}  
is future-pointing in the untrapped region, where we used $\varepsilon_{uv}=\exp(-f)$ and $\varepsilon^{uv}=-\exp(f)$.

In the double null coordinates, the quasi-local mass $m_{\rm L}$ is expressed as
\begin{align}
\label{qlm2}
m_{\rm L} &= \frac{(n-2)V_{n-2}^{(k)}}{2\kappa_n^2}\sum_{p=0}^{[n/2]}
{\tilde \alpha}_{(p)}r^{n-1-2p}\left(k+\frac{2r^2e^{f}}{(n-2)^2}\theta_{+}\theta_{-}\right)^p,
\end{align}  
while the unified first law (\ref{1stlaw1}) is written as
\begin{align}
\partial_{v}m_{{\rm L}}&=
\frac{1}{n-2}V_{n-2}^{(k)}e^fr^{n-1}(T_{uv}\theta_+-T_{vv}\theta_-), \label{m_v} \\
\partial_{u}m_{{\rm L}}&=
\frac{1}{n-2}V_{n-2}^{(k)}e^fr^{n-1}(T_{uv}\theta_- -T_{uu}\theta_+). \label{m_u} 
\end{align}  
%
The null energy condition for the matter field implies
\begin{align}
T_{uu}\ge 0,~~~T_{vv} \ge 0, \label{nec}
\end{align}
while the dominant energy condition implies
\begin{align}
T_{uu} \ge 0,~~T_{vv}\ge 0,~~T_{uv}\ge 0. \label{dec}
\end{align}  
The quasi-local mass $m_{\rm L}$ has the following monotonic property independent of the signs of the Lovelock coupling constants ${\alpha}_{(p)}$. 
\begin{Prop}
\label{th:monotonicity}
({\it Monotonicity.}) 
Under the dominant energy condition, 
$m_{\rm L}$ is non-decreasing (non-increasing) in any outgoing  
(ingoing) spacelike or null direction on an untrapped surface.
\end{Prop}

Since the equations (\ref{m_v}) and (\ref{m_u}) do not contain the Lovelock coupling constants explicitly, the proof of the above proposition is the same as the Proposition 4 in~\cite{mn2008}.

Next, we show the positivity of $m_{\rm L}$ in the spherically symmetric spacetime ($k=1$).
\begin{Prop}
\label{th:positivity}
({\it Positivity.}) 
Suppose $k=1$.
If the dominant energy condition holds on an untrapped spacelike hypersurface with a regular center in spherically symmetric spacetime, then $m_{\rm L}\ge 0$.
\end{Prop}
\noindent
{\it Proof}.
The point where $r=0$ is called {\it center} if it defines the boundary of $(M^2, g_{ab})$.
A central point is called {\it regular} if 
\begin{equation}
\label{r-center}
\frac{2}{(n-2)^2}e^f r^2\theta _+\theta _- +1 \simeq \eta r^2
\end{equation}  
holds around the center and {\it singular} otherwise, 
where a constant $\eta $ is assumed to be non-zero.
($\eta=0$ and $m_{\rm L}\equiv 0$ are satisfied for Minkowski, which has a regular center.)
Hence, the regular center is surrounded by untrapped surfaces.
From Eq.~(\ref{qlm2}), we obtain
\begin{align}
\label{qlm-center}
m_{\rm L} &\simeq \frac{V_{n-2}^{(k)}(n-2)!r^{n-1}}{2\kappa_n^2}\sum_{p=0}^{[n/2]}
\frac{\alpha_{(p)}\eta^p}{(n-1-2p)!}
\end{align}  
around the regular center.
From Eq.~(\ref{qlm-center}), we obtain
\begin{align}
\partial_{v}m_{{\rm L}} &\simeq \frac{(n-1)(n-3)!V_{n-2}^{(k)}}{2\kappa_n^2}
r^{n-1}\theta_{+}\sum_{p=0}^{[n/2]}\frac{\alpha_{(p)}\eta^p}{(n-1-2p)!},
 \label{m_v-c} \\
\partial_{u}m_{{\rm L}} &\simeq \frac{(n-1)(n-3)!V_{n-2}^{(k)}}{2\kappa_n^2}
r^{n-1}\theta_{-}\sum_{p=0}^{[n/2]}\frac{\alpha_{(p)}\eta^p}{(n-1-2p)!} \label{m_u-c}
\end{align}  
around the regular center.
By Eqs.~(\ref{m_v-c}) and (\ref{m_u-c}) and Proposition~\ref{th:monotonicity}, the inequality $\sum_{p=0}^{[n/2]}(\alpha_{(p)}\eta^p)/(n-1-2p)! \ge 0$ is satisfied under the dominant energy condition since the regular center is surrounded by untrapped surfaces.
As a result, by Eq.~(\ref{qlm-center}), $m_{\rm L}$ is non-negative around the regular center.
Then, the proposition follows from Proposition~\ref{th:monotonicity}.
\qed

\bigskip

Proposition~\ref{th:positivity} shows the positivity of $m_{\rm L}$ in the untrapped region with a regular center.
On the other hand, we may obtain a positive lower bound for $m_{\rm L}$ if there is a marginal surface.
The quasi-local mass on the marginal surface is given by 
\begin{align}
\label{qlm-horizon}
m_{\rm h} &= \frac{(n-2)V_{n-2}^{(k)}}{2\kappa_n^2}\sum_{p=0}^{[n/2]}{\tilde\alpha}_{(p)}r_{\rm h}^{n-1-2p}k^p, 
\end{align}  
where $r_{\rm h}:=r|_{\theta_+\theta_-=0}(>0)$.
Using the Proposition \ref{th:monotonicity}, the following mass inequality is shown~\cite{error};
\begin{Prop}
\label{th:mass}
({\it Mass inequality.}) 
If the dominant energy condition holds, then $m_{\rm L} \ge m_{\rm h}$ holds on an untrapped spacelike hypersurface of which the inner boundary is a marginal surface with radius $r_{\rm h}$.
For ${\alpha}_{(p)}k^p \ge 0$ and $\sum_{p=0}^{[n/2]}{\tilde\alpha}_{(p)}k^p \ne 0$, this lower bound $m_{\rm h}$ is positive for $k=1,-1$ and non-negative for $k=0$.
\end{Prop}

\bigskip

Using Propositions~\ref{th:asymptotics}, \ref{th:positivity}, and \ref{th:mass}, we can show the following positive mass theorem in asymptotically flat spherically symmetric spacetime in Lovelock gravity.
\begin{Prop}
\label{coro:positivemass}
({\it Positivity of the ADM mass.}) 
Suppose ${\alpha}_{(p)}\ge 0$ and $\sum_{p=0}^{[n/2]}{\tilde\alpha}_{(p)} \ne 0$ and the dominant energy condition is satisfied in a spherically symmetric, asymptotically flat, and regular spacetime.
Then, the ADM mass is non-negative.
\end{Prop}
\noindent
{\it Proof}.
First let us consider the case where there is no marginal surface in the spacetime.
If there is a regular center, the ADM mass is non-negative by Proposition~\ref{th:positivity}.
If there is no regular center, there is at least one wormhole throat on a spacelike hypersurface, where
\begin{align}
\label{throat}
\zeta^\mu \nabla_\mu {\cal A} =0
\end{align}  
is satisfied with a radial spacelike vector $\zeta^\mu$.
This equation gives
\begin{align}
\zeta^u\theta_-+\zeta^v \theta_+ =0.
\end{align}  
Since $\zeta^u\zeta^v<0$, $\theta_+=\theta_-=0$ or $\theta_+\theta_->0$ is satisfied.
Because the asymptotically flat region consists of untrapped surfaces, there is at least one marginal surface on a spacelike hypersurface by the mean value theorem in the latter case.
Hence, both of them reduce to the case with a marginal surface on a spacelike hypersurface.
If there is a marginal surface on a spacelike hypersurface, the ADM mass is positive because of Proposition~\ref{th:mass} and the fact that the asymptotically flat region consists of untrapped surfaces.

\qed

\bigskip

In addition to the monotonicity and positivity, $m_{\rm L}$ has the following rigidity property.
This proposition is shown for non-negative Lovelock coupling constants and claims that if the generalized Misner-Sharp mass is vanishing somewhere, the spacetime is locally vacuum.
(The proof is similar to Proposition 1 in~\cite{maeda2008}.)
\begin{Prop}
\label{lm:zeromass}
({\it Rigidity.}) 
Under the dominant energy condition with $k=1$, $\alpha_{(p)} \ge 0$, and $\sum_{p=1}^{[n/2]}{\tilde\alpha}_{(p)} \ne 0$, if $m_{\rm L} \equiv 0$ is satisfied, then $T_{\mu\nu}\equiv  0$ holds there.
\end{Prop}

\bigskip

We have shown that $m_{\rm L}$ is a well-defined quasi-local mass and a natural generalization of the Misner-Sharp mass in general relativity.
In the next section, we use $m_{\rm L}$ to evaluate the mass of a dynamical black hole.

\subsection{Branches of solutions}
\label{sec:branch}
Actually, the Lovelock equations admit multiple branches of solutions.
In order to see this, we define the following function:
\begin{align}
W(x):=&\frac{2\kappa_n^2m_{\rm L}}{(n-2)V_{n-2}^{(k)}}+ \sum_{p=0}^{[n/2]}{\tilde \alpha}_{(p)}r^{n-1-2p}x^p.\label{W}
\end{align}  
$W(x)=0$ with $x=k-(Dr)^2$ is equivalent to Eq.~(\ref{qlm}) and consistent with the Lovelock equations.
Here we assume $(Dr)^2\ne 0$.
The function $W(x)$ is regarded as a function of $x$ with coefficients depending on $r$.
Hence, if $W(x)=0$ has $N(\le [(n-1)/2])$ real solutions of $x$ for a given value of $r$, there are $N$ real branches of solutions there.

If we have $(\lim_{x\to -\infty} W(x))(\lim_{x\to +\infty} W(x))<0$, there is at least one real solution for $W(x)=0$ independent of the Lovelock coupling constants by the mean value theorem.
Hence, the following proposition is shown.
\begin{Prop}
\label{lm:odd2}
({\it Existence of real solutions.}) 
Let $s$ be the largest integer with non-zero ${\tilde \alpha}_{(s)}$ in $1 \le s \le [n/2]$ and suppose $(Dr)^2\ne 0$.
Then, the field equations admit real metric for any given value of $r$ if $s$ is odd.
\end{Prop}


Indeed, in the second-order Lovelock gravity ($s=2$), namely in Einstein-Gauss-Bonnet gravity, the metric of the Boulware-Deser-Wheeler solution with negative mass can be complex for some value of $r$~\cite{bdw,tm2005}.
The above proposition claims that at least one of three branches is real in the cubic Lovelock vacuum solution ($s=3$)~\cite{ds2005}.
We should note that although there is a real branch for each value of $r$, it is not necessarily true that the same branch would be real for all values of $r$.
Therefore, we cannot guarantee that there exists a single spacetime which extends over all values of $r$.
In general Lovelock gravity, the term with the highest power in $W$ is $x^{(n-1)/2}$ in odd dimensions, while it is $x^{(n-2)/2}$ in even dimensions (since ${\tilde \alpha}_{(n/2)}\equiv 0$).
Thus, the following corollary is given from Proposition~\ref{lm:odd2}.
\begin{Coro}
\label{coro:lovelock}
In general Lovelock gravity in $n(\ge 4)$ dimensions with $(Dr)^2\ne 0$, the field equations admit a real metric for any given value of $r$ if $n=4q$ or $4q+3$, where $q$ is an integer.
\end{Coro}


In contrast, in the case where $n=4q+1$ or $4q+2$, $W(x)=0$ may not have a real solution for some $r$ and there the metric becomes complex and unphysical.
We will see this in the Schwarzschild-Tangherlini-type solution in Proposition~\ref{th:r=0}.

To close this section, we comment that the form of the quasi-local mass $m_{\rm L}$ becomes quite simple in the case where there is a unique maximally symmetric solution~\cite{DCBH,DCBH2}.
The analysis in such a case is presented in Appendix B.

\section{Dynamical black holes}
\label{sec4}
In this section, we study the properties of dynamical black holes defined by a future outer trapping horizon.
Trapping horizons were defined by Hayward~\cite{hayward1994,hayward1996}. 
Our system contains those in~\cite{ck2005,ac2007,cc2007,cchk2008} as special cases with a particular type of matter field. 
In this section, we don't use the orientation in the previous section.
Instead, we set $\theta_+=0$ on the marginal surface without loss of generality.
A marginal surface is classified as follows.
\begin{dn}
\label{def:4-msphere}
A marginal surface is {\it future} if $\theta_-<0$, {\it past} if
$\theta_{-}>0$, {\it bifurcating} if $\theta_-=0$, {\it outer} if
$\partial_{u}\theta_{+}<0$, {\it inner} if $\partial_{u}\theta_{+}>0$ and {\it
degenerate} if $\partial_{u}\theta_{+}=0$.
\end{dn}
A future (past) marginal surface means that the outgoing (ingoing) null rays are marginally trapped, namely instantaneously parallel on the horizon, on the marginal surface.
A Kodama vector (\ref{kodamavector}) is future-pointing (past-pointing) on the future (past) marginal surface.
\begin{dn}
\label{def:t-horizon}
A {\it trapping horizon} is the closure of a hypersurface 
foliated by future or past, outer or inner marginal surfaces.
\end{dn}
Among all the classes, a {\it future outer} trapping horizon defines a dynamical black hole because it means that the ingoing null rays are converging ($\theta_-<0$), while the outgoing null rays are instantaneously parallel on the horizon ($\theta_+=0$) and diverging just outside the horizon and converging just inside ($\partial_{u}\theta_{+}<0$)~\cite{hayward1994,hayward1996}.
On the other hand, a {\it past outer} trapping horizon defines a dynamical white hole.
Accordingly, we call the closure of a hypersurface foliated by bifurcating or degenerate marginal surfaces, a {\it bifurcating trapping horizon} or a {\it degenerate trapping horizon}, respectively.

\subsection{Properties of trapping horizon}
In this subsection, we show some basic properties of trapping horizons.
The following lemma is essential in the proofs of the later propositions. 
\begin{lm}
\label{lm:+v}
Under the null energy condition with ${\alpha}_{(p)}k^{p-1} \ge 0$ for $p\ge 1$ and $\sum_{p=1}^{[n/2]}{\tilde\alpha}_{(p)}k^{p-1} \ne 0$, $\partial_{v}\theta_{+}\le 0$ on the trapping horizon.
\end{lm}
{\it Proof}. 
Evaluating the $(v,v)$ component of (\ref{comp}) on the trapping horizon, we obtain~\cite{typo}
\begin{align}
-\kappa_n^2T_{vv} = \sum_{p=0}^{[n/2]}\frac{p{\tilde\alpha}_{(p)}k^{p-1}}{r_{\rm h}^{2(p-1)}}\partial_{v}\theta_{+}.\label{Tvv}
\end{align}
The lemma follows from the above equation.
\qed

\bigskip

It is noted that, in Einstein-Gauss-Bonnet gravity, the inequality $\partial_{v}\theta_{+}\le 0$ was shown independent of the signs of the coupling constants and $k$~\cite{nm2008}.
This nice result is achieved by classifying the solutions in an appropriate manner.
There the solution can be classified into two branches, one has the general relativistic limit (GR branch), while the other does not (non-GR branch). 
Then, it was successfully shown that the trapping horizon in the GR branch has the same properties as the general relativistic case, while the properties of the non-GR-branch solutions are pathological such as the decreasing area of a black hole~\cite{mn2008}.
The key step for this simple result is to clarify the domain of existence for the trapping horizon depending on the branches and the parameters.
(See Proposition 4 in~\cite{mn2008}.)
This is possible in Einstein-Gauss-Bonnet gravity because the definition of the quasi-local mass (\ref{qlm}) has a quadratic form, which allows us to solve for $(Dr)^2$.
However, in general Lovelock gravity, it is a very complicated task to distinguish the branches of solutions.
Therefore, we focus in this section mainly on the case with ${\alpha}_{(p)}k^{p-1}\ge 0~(p\ge 1)$ and $\sum_{p=1}^{[n/2]}{\tilde\alpha}_{(p)}k^{p-1} \ne 0$, in which the analysis is drastically simplified because of Lemma~\ref{lm:+v}.

Now let us clarify the basic properties of the trapping horizon.
Since the trapping horizon is foliated by marginal surfaces, 
\begin{align}
{\cal L}_\xi \theta_{+}=\xi^v\partial_v\theta_{+}+\xi^u\partial_u\theta_{+}=0 \label{expansionderiv}
\end{align}
on the trapping horizon, where $\xi^\mu$ is the generator of the trapping horizon.
The following lemma will be also used in the proofs of the later propositions. 
\begin{lm}
\label{lm:2}
$\xi^u=0$ and $\partial_v\theta _{+}=0$ on a null non-degenerate trapping horizon.
\end{lm}
{\it Proof}. 
If the trapping horizon is null, there are two possibilities for its generator, $\xi^u=0$ or $\xi^v=0$.
In the case of $\xi^v=0$, Eq.~(\ref{expansionderiv}) gives $\partial_u\theta_{+}=0$ and hence it is the degenerate-type.~\footnote{Such a null degenerate trapping horizon is realized in the Friedmann-Robertson-Walker spacetime $ds^2=-dt^2+(t/t_0)\{d\chi^2+\chi^2(d\theta^2+\sin^2\theta d\phi^2)\}$.}
Therefore, $\xi^u=0$ for a null and non-degenerate trapping horizon.~\footnote{The trapping horizon in the Schwarzschild-Tangherlini spacetime is an example.}
In this case, Eq.~(\ref{expansionderiv}) gives $\partial_v\theta _{+}=0$ on the trapping horizon.
\qed

\bigskip

Using Lemmas~\ref{lm:+v} and \ref{lm:2}, the following propositions are shown.
They mean that null or causal observers cannot escape from the trapped region by crossing a future outer trapping horizon, which suits the concept of a black-hole horizon as a one-way membrane.
\begin{Prop}
({\it Signature law.})
\label{prop:sig} 
Under the conditions in Lemma~\ref{lm:+v}, an outer (inner) 
trapping horizon is non-timelike (non-spacelike).
\end{Prop}
\vspace{5mm}
{\it Proof}. 
From Eq. (\ref{expansionderiv}), we have  
\begin{eqnarray} 
\xi ^u=-\frac{\partial_v\theta_{+}}{\partial_u\theta_{+}}\xi ^v,
\end{eqnarray}
on the non-degenerate trapping horizon.
Thus, by Lemmas~\ref{lm:+v} and \ref{lm:2}, $\xi ^v\xi ^u \le (\ge)0 $ on the outer (inner) trapping horizon. 

 \qed

\bigskip

\begin{Prop}
({\it Trapped side.})
\label{prop:futureinner}
Under the conditions in Lemma~\ref{lm:+v}, the outside (inside) region of an inner trapping horizon is trapped (untrapped). 
While the future (past) domain of a future outer trapping horizon is trapped (untrapped), the future (past) domain of a past outer trapping horizon is untrapped (trapped).
\end{Prop}
\noindent
{\it Proof}. 
Since we adopted the symmetric slice to define a trapped surface, we consider a vector $s^\mu(\partial/\partial x^\mu)=s^v(\partial/\partial v)+s^u(\partial/\partial u)$ to prove the proposition.
Along the vector $s^\mu$, we obtain 
${\cal L}_s (\theta_{+}\theta_{-})=\theta_{-}(s^v\partial_v\theta_{+}+s^u\partial_u\theta_{+})$ 
on a trapping horizon.
First, let us take $s^\mu$ to be an outgoing spatial vector.
Since we obtain
\begin{align}
{\ma L}_s {\cal A}=V_{n-2}^{(k)}r_{\rm h}^{n-2}\theta_{-} s^u \label{lieS}
\end{align}
on the trapping horizon, $s^v>(<)0$ and $s^u<(>)0$ on the future (past) trapping horizon in order for $s^\mu$ to be outward-pointing.
By Proposition~\ref{prop:sig}, an inner trapping horizon is non-spacelike.
Then, by Lemma~\ref{lm:+v}, ${\cal L}_s (\theta_{+}\theta_{-})>0$ holds on a future inner or past inner trapping horizon, which means that outside the trapping horizon is in the trapped region.
Next let $s^\mu$ be a future-directed timelike vector, where $s^v>0$ and $s^u>0$.
Then, by Lemma~\ref{lm:+v}, ${\cal L}_s (\theta_{+}\theta_{-})>(<)0$ holds on a future (past) outer trapping horizon.
This means that future (past) domain of the future (past) outer trapping horizon is in the trapped region.

\qed

\bigskip

Next we show the following area law for a dynamical black hole.
\begin{Prop}
\label{arealaw}
({\it Area law.}) 
Under the conditions in Lemma~\ref{lm:+v}, the area of a future (past) inner trapping horizon is non-increasing (non-decreasing) into the future.
The area of an outer trapping horizon is non-decreasing along its generator.
\end{Prop}
\noindent
{\it Proof}. 
Along the generator of the trapping horizon $\xi^\mu$, we obtain
\begin{align}
{\ma L}_\xi {\cal A}=V_{n-2}^{(k)}r_{\rm h}^{n-2}\theta_{-} \xi^u.\label{lieA}
\end{align}
Since the generator of a null non-degenerate trapping horizon satisfies $\xi^u = 0$ by Lemma~\ref{lm:2}, we can take $\xi^v >0$ for a non-degenerate trapping horizon without loss of generality.
The area of a null non-degenerate trapping horizon ($\xi^u = 0$) is constant along its generator. 
For a timelike trapping horizon, we set $\xi^u > 0$ in order for the generator to be future-pointing.
Equation (\ref{lieA}) shows that the area of a timelike future (past) trapping horizon is decreasing (increasing).
It also shows that the generator of a spacelike future (past) trapping horizon is outward-pointing if $\xi^u<(>)0$.
Finally, the proposition follows from Proposition~\ref{prop:sig}.
\qed

\bigskip

In the case where the generator is spacelike, the above proposition simply means that the area of the trapping horizon increases in the outward direction.
Here it should be emphasized that a future outer trapping horizon can be timelike and its area can decrease in some parameter space, as shown in Einstein-Gauss-Bonnet gravity~\cite{nm2008,maeda2010}.
There, the non-increasing area of a black hole is realized only in the non-GR branch of solutions.

Finally, we derive the dynamical entropy of Lovelock black holes.
For this purpose, we first define the dynamical surface gravity and temperature of a dynamical black hole in a geometrical way using a Kodama vector~(\ref{kodamavector}) in a similar manner as a timelike Killing vector is used in the stationary case.
This is because a Kodama vector satisfies Eq.~(\ref{ksquare}) and hence it provides a preferred time direction in the untrapped region. 
It is shown that $K^a$ is null and given by $K_a=D_a r$ on the trapping horizon.
Hence, in a similar manner to the surface gravity in the stationary spacetime, we define the dynamical surface gravity $\kappa _{\rm TH}$ by
\begin{align}
K^bD_{[b}K_{a]}=\pm \kappa _{\rm TH}K_a,
\end{align}
evaluated on the trapping horizon~\cite{hayward1998}.
In the above definition, $K^a$ is assumed to be future-pointing and the sign in the right-hand side is chosen such that $\kappa _{\rm TH}$ is positive (negative) for an outer (inner) trapping horizon.
It is noted that the Kodama vector (\ref{kodamavector}) is future-pointing (past-pointing) for a future (past) trapping horizon because we set $\theta_+=0$ on the marginal surface. (See the expression~(\ref{kodama2}).)
As a result, we obtain
\begin{align}
\kappa_{\rm TH} =\frac12 D^2 r\biggl|_{r=r_{\rm h}}.\label{surfacegravity}
\end{align}
(See Section 3.3 in~\cite{nm2008} for the derivation.)
Since we have
\begin{align}
D^2 r=-2e^{f}\partial_u\partial_v r,
\end{align}
$D^2 r>(<)0$ on an outer (inner) trapping horizon, while $D^2 r=0$ for a degenerate trapping horizon.
As in the stationary spacetime, we define the temperature of a trapping horizon by $T_{\rm TH}:=\kappa_{\rm TH}/(2\pi)$.

The unified first law (\ref{1stlaw1}) can be written as
\begin{align}
{\cal A}\psi_a=&r^{n-3}D_a \biggl[\frac{m_{\rm L}}{r^{n-3}}-\frac{(n-2)V_{n-2}^{(k)}}{2\kappa_n^2}\sum_{p=0}^{[n/2]}{\tilde \alpha}_{(p)}r^{2-2p}k^p\biggl] \nonumber \\
&+\frac{(n-2)V_{n-2}^{(k)}r^{n-2}}{2\kappa_n^2}\sum_{p=0}^{[n/2]}{\tilde \alpha}_{(p)}\biggl[(2-2p)\biggl(\frac{k}{r^2}\biggl)^p-(2-2p)\left(  \frac{k- (Dr)^2}{r^2} \right)^{p}\biggl]D_ar \nonumber \\
&+\frac{(n-2)V_{n-2}^{(k)}r^{n-2}}{2\kappa_n^2} \sum_{p=0}^{[n/2]}{\tilde \alpha}_{(p)}\biggl[p \frac{D^2r}{r}
\left(  \frac{k-(Dr)^2 }{r^2} \right)^{p-1}\biggl]D_a r.
\end{align}
The first and second lines in the right-hand side vanish on a trapping horizon.
Evaluating the above equation on a trapping horizon with its generator $\xi^a$, we obtain  
\begin{align}
{\cal A}\psi_a\xi^a=&\frac{\kappa_{\rm TH}}{\kappa_n^2}\xi^aD_a \biggl(\sum_{p=0}^{[n/2]}{\tilde \alpha}_{(p)}\frac{p(n-2)}{n-2p} k^{p-1}V_{n-2}^{(k)}r^{n-2p} \biggl).
\end{align}
Identifying the right-hand side as $T_{\rm TH}{\ma L}_\xi S_{\rm TH}$, we derive the dynamical entropy $S_{\rm TH}$ as
\begin{align}
S_{\rm TH}:=&\frac{2\pi}{\kappa_n^2}\sum_{p=0}^{[n/2]}{\tilde \alpha}_{(p)}\frac{p(n-2)}{n-2p} k^{p-1}V_{n-2}^{(k)}r_{\rm h}^{n-2p} , \label{S} \nonumber \\
=&\frac{{\tilde \alpha}_{(1)}{\cal A}_{\rm h}}{4G_n}+\frac{2\pi}{\kappa_n^2}\sum_{p=2}^{[n/2]}{\tilde \alpha}_{(p)}\frac{p(n-2)}{n-2p} k^{p-1}V_{n-2}^{(k)}r_{\rm h}^{n-2p},
\end{align}
where we set the integration constant to zero in order to be compatible with the Wald entropy in the static case which will be derived in Section~\ref{subsec:thermo}.

The variation of $S_{\rm TH}$ along the generator of a trapping horizon is given by
\begin{align}
{\ma L}_\xi S_{\rm TH}=&\frac{2\pi}{\kappa_n^2} \sum_{p=0}^{[n/2]}{\tilde \alpha}_{(p)}p k^{p-1}V_{n-2}^{(k)}r_{\rm h}^{n-2p}\theta_{-} \xi^u
\end{align}
and the unified first law (\ref{1stlaw1}) gives
\begin{align}
{\ma L}_\xi m_{\rm L}=T_{\rm TH}{\ma L}_\xi S_{\rm TH}+P_{\rm TH}{\ma L}_\xi V,
\end{align}
where $P_{\rm TH}:=P|_{r=r_{\rm h}}$.
Then, in a similar manner to Proposition~\ref{arealaw}, the following entropy law is shown.
\begin{Prop}
\label{entropylaw}
({\it Entropy law.}) 
Under the conditions in Lemma~\ref{lm:+v}, the entropy of a future (past) inner trapping horizon is non-increasing (non-decreasing) into the future.
The entropy of an outer trapping horizon is non-decreasing along its generator.
\end{Prop}

\bigskip

It is noted that, in Einstein-Gauss-Bonnet gravity, while a future outer trapping horizon can be timelike and its area can decrease in the non-GR branch, its dynamical entropy still increases~\cite{nm2008,maeda2010}.
This supports the intuition that the entropy law is more fundamental than the area law in black-hole physics.

\subsection{Exact solutions with matter}
\label{ex-matter}
We have seen some of the basic properties of dynamical black holes in Lovelock gravity.
In this subsection, we present several exact solutions with various kinds of matter fields as concrete examples of dynamical spacetimes with trapping horizons.
It also shows the usefulness of the concept of the generalized Misner-Sharp mass, which is the first integral of the Lovelock equations, to obtain exact solutions.

First let us consider the following energy-momentum tensor;
\begin{eqnarray}
{T}^\mu_{~~\nu}=\mbox{diag}(-\mu(y),P_{\rm r}(y),P_{\rm t}(y),\cdots,P_{\rm t}(y))
\end{eqnarray}
in the following diagonal coordinates;
\begin{equation}
ds^2=-e^{2\Phi(t,\rho)}dt^2+e^{2\Psi(t,\rho)}d\rho^2+r(t,\rho)^2\gamma_{ij}dz^i dz^j.
\end{equation}  
$\mu$, $P_{\rm r}$, and $P_{\rm t}$ are interpreted as the energy density, radial pressure, and tangential pressure, respectively.
We obtain the basic equations as
\begin{eqnarray}
{P_{\rm r}}'&=& -(\mu+P_{\rm r})\Phi'-(n-2)(P_{\rm r}-P_{\rm t})\frac{r'}{r}, \label{basic1-ap}\\
\dot{\mu}&=& -(\mu+P_{\rm r})\dot{\Psi}-(n-2)(\mu+P_{\rm t})\frac{\dot{r}}{r}, \label{basic2-ap}\\
m_{\rm L}' &=& V_{n-2}^{(k)}\mu r' r^{n-2},   \label{basic3-ap}\\
\dot{m}_{\rm L} &=& -V_{n-2}^{(k)}P_{\rm r} \dot{r} r^{n-2}, \label{basic4-ap}\\ 
0&=&(-\dot{r'}+\Phi' \dot{r}+\dot{\Psi}r')\sum_{p=0}^{[n/2]}{\tilde\alpha}_{(p)}p r^{2(1-p)}(k+e^{-2\Phi}\dot{r}^2-e^{-2\Psi}{r'}^2)^{p-1},\label{basic5-ap}\\
m_{\rm L} &=& \frac{(n-2)V_{n-2}^{(k)}}{2\kappa_n^2}\sum_{p=0}^{[n/2]}{\tilde \alpha}_{(p)}r^{n-1-2p}(k+e^{-2\Phi}\dot{r}^2-e^{-2\Psi}{r'}^2)^p,\label{basic6-ap}
\end{eqnarray}  
where a dot and a prime denote the differentiation with respect to $t$ and $\rho$, respectively. 
Equations (\ref{basic1-ap}) and (\ref{basic2-ap}) are given from the conservation equation $\nabla_\mu T^\mu_{~\nu}=0$, while Eqs.~(\ref{basic3-ap}) and (\ref{basic4-ap}) are given from the unified first law (\ref{1stlaw1}).
Equation (\ref{basic5-ap}) is the $(t,\rho)$ component of the field equation, while Eq.~(\ref{basic6-ap}) comes from the definition of the generalized Misner-Sharp mass.
Equation~(\ref{besij}) gives an auxiliary equation.

Here we need a comment on Eq.~(\ref{basic5-ap}). 
The field equation~(\ref{beq-simp}) gives
\begin{align}
\kappa_n^2\biggl(T^{a}_{~b}- \frac{1}{2} \delta^a_{~b}T^{d}_{~d}\biggl) = & -(n-2)r^{-1}\biggl(D^aD_br-\frac12 (D^2r)\delta^a_{~b}\biggl)\sum_{p=0}^{[n/2]}{\tilde\alpha}_{(p)}p\left(  \frac{k-(Dr)^2 }{r^2} \right)^{p-1}.\label{exact-condition}
\end{align}
The $(t,t)$, $(t,\rho)$, and $(\rho,\rho)$ components of the left-hand side are $-\kappa_n^2(\mu+P_{\rm r})/2$, $0$, and $\kappa_n^2(\mu+P_{\rm r})/2$, respectively.
The $(t,\rho)$ component of Eq.~(\ref{exact-condition}) exactly gives Eq.~(\ref{basic5-ap}). 
It is seen that there are two possibilities to solve the $(t,\rho)$ component; $D^tD_\rho r=0$ and 
\begin{align}
\sum_{p=0}^{[n/2]}{\tilde\alpha}_{(p)}p\left(  \frac{k-(Dr)^2 }{r^2} \right)^{p-1}=0.\label{special-c}
\end{align}
Since the condition (\ref{special-c}) vanishes the right-hand side of Eq.~(\ref{exact-condition}), it gives a constraint on the matter field $P_{\rm r}+\mu=0$.
Therefore, under an assumption $P_{\rm r}+\mu \ne 0$, the $(t,\rho)$ component of Eq.~(\ref{exact-condition}) gives $D^tD_\rho r=0$, namely Eq.~(\ref{basic5-ap}) reduces to 
\begin{eqnarray}
0=-\dot{r'}+\Phi' \dot{r}+\dot{\Psi}r'.
\end{eqnarray}  
It is noted that the condition (\ref{special-c}) does not have a general relativistic counterpart.

\subsubsection{Friedmann-Robertson-Walker solution with a perfect fluid}
Let us consider the case with a perfect fluid, where $P_{\rm r}=P_{\rm t}=:{\bar P}$, for the Friedmann-Robertson-Walker metric ($k=1$);
\begin{equation}
ds^2=-dt^2+a(t)^2\biggl(\frac{d\rho^2}{1-K\rho^2}+\rho^2\gamma_{ij}dz^i dz^j\biggl),
\end{equation}  
where $K=\pm 1,0$ is the spatial curvature of the $(n-1)$-dimensional space. 
With an equation of state ${\bar P}=(\gamma-1)\mu$, the field equations reduce to the following master equation for $a(t)$;
\begin{eqnarray}
\frac{2\kappa_n^2\mu_0}{(n-1)(n-2)}  = a^{(n-1)\gamma}\sum_{p=0}^{[n/2]}{\tilde \alpha}_{(p)}\biggl(\frac{\dot{a}^2}{a^2}+\frac{K}{a^2}\biggl)^p. \label{master-FRW}
\end{eqnarray}  
This is the generalized Friedmann equation and $\mu$ and $m_{\rm L}$ are given as
\begin{eqnarray}
\mu(t)&=&\mu_0 a^{-(n-1)\gamma}, \\
m_{\rm L}(t,\rho) &=& \frac{1}{n-1}V_{n-2}^{(1)}\mu_0 a^{(n-1)(1-\gamma)}\rho^{n-1},
\end{eqnarray}  
where $\mu_0$ is a constant.
The trapping horizon in this spacetime is represented by 
\begin{eqnarray}
1=\rho^2a^2\biggl(\frac{\dot{a}^2}{a^2}+\frac{K}{a^2}\biggl).
\end{eqnarray}  
The dynamics of this spacetime was fully investigated in~\cite{df1990}.
The thermodynamical properties of the trapping horizon in this spacetime was investigated in~\cite{ck2005,ac2007,cc2007}.

For the dust case ($\gamma=1$), there is an exact solution with $a(t)=a_0$, where $a_0$ is determined algebraically by
\begin{align}
0=- \sum_{p=0}^{[n/2]}{\tilde \alpha}_{(p)}(n-1-2p)\left(  \frac{K}{a_0^2} \right)^{p}.
\end{align}
The parameters are fixed to give real $a_0$.
The energy density of dust is constant and given by 
\begin{eqnarray}
\mu = \frac{(n-1)(n-2)}{2\kappa_n^2}\sum_{p=0}^{[n/2]}{\tilde \alpha}_{(p)}\biggl(\frac{K}{a_0^2}\biggl)^p. \label{Einsteinstatic}
\end{eqnarray}    
This is the generalized Einstein static universe in Lovelock gravity~\cite{df1990}, where $K=0$ gives ${\tilde \alpha}_{(0)}=0$ and the spacetime is Minkowski.

\subsubsection{Generalized Lema{\^ i}tre-Tolman-Bondi solution with a dust fluid}
The Friedmann-Robertson-Walker spacetime is a homogeneous spacetime.
In a general inhomogeneous case, it is difficult to make the system reduce to a single master equation.
However, it is possible in the case with a timelike dust fluid ($P_{\rm r}=P_{\rm t}\equiv 0$).
Then, the system reduces to the following metric:
\begin{equation}
ds^2=-dt^2+\frac{{r'}^2}{k+h(\rho)}d\rho^2+r(t,\rho)^2\gamma_{ij}dz^i dz^j
\end{equation}  
with a master equation for $r$;
\begin{eqnarray}
m_{\rm L}(\rho)= \frac{(n-2)V_{n-2}^{(k)}}{2\kappa_n^2}\sum_{p=0}^{[n/2]}{\tilde \alpha}_{(p)}r^{n-1-2p}(\dot{r}^2-h(\rho))^p,\label{master-LTB}
\end{eqnarray}  
where the generalized Misner-Sharp mass $m_{\rm L}(\rho)$ and $h(\rho)$ are arbitrary functions of $\rho$.
The energy density of dust is given by 
\begin{eqnarray}
\mu(t,\rho) = \frac{m_{\rm L}'}{V_{n-2}^{(k)} r' r^{n-2}}.
\end{eqnarray}  
The trapping horizon in this spacetime is represented by 
\begin{eqnarray}
\dot{r}^2-h(\rho)=k.
\end{eqnarray}  
This is the generalized Lema{\^ i}tre-Tolman-Bondi solution in Lovelock gravity.
The solution with $k=1$ was recently obtained and investigated in the context of gravitational collapse~\cite{osj2011}.

\subsubsection{Generalized Vaidya solution with a null dust fluid}
A null dust fluid is another kind of matter with which we easily obtain an exact solution.
The energy-momentum tensor for a null dust is given by
\begin{eqnarray}
{T}_{\mu\nu}=\mu l_{\mu}l_{\nu},
\end{eqnarray}
where $\mu$ is the energy density and $l_\mu$ is a null vector.
With this matter field, there is an exact solution;
\begin{align}
ds^2=&-f(v,r)dv^2+2dvdr+r^2\gamma_{ij}dz^i dz^j,\\
l_{\mu}=&-\partial_{\mu} v,
\end{align}  
where $f(v,r)$ is given algebraically by the following master equation;
\begin{align}
m_{\rm L}(v) = \frac{(n-2)V_{n-2}^{(k)}}{2\kappa_n^2}\sum_{p=0}^{[n/2]}{\tilde \alpha}_{(p)}r^{n-1-2p}(k-f)^p,
\end{align}  
where the generalized Misner-Sharp mass $m_{\rm L}$ is an arbitrary function of the advanced time $v$.
The energy density $\mu$ is given by
\begin{align}
\mu(v,r)=\frac{n-2}{2\kappa_n^2r^{n-2}}\frac{\partial m_{\rm L}}{\partial v}.
\end{align}  
This is the generalized Vaidya solution and reduces to the Schwarzschild-Tangherlini-type vacuum solution if $m_{\rm L}$ is constant, which is investigated in the following sections.
This is a dynamical and inhomogeneous spacetime and the trapping horizon is given by $f(v,r)=0$.
The solution with $k=1$ was obtained and used to discuss the thermodynamical properties of the trapping horizon in~\cite{cchk2008}.
Also the solution was obtained in dimensionally continued gravity with arbitrary $k$ and investigated in the context of gravitational collapse~\cite{vaidya-L}.

\section{Vacuum solutions}
\label{sec5}
In this section, we study vacuum solutions in more detail.
The unified first law (\ref{1stlaw1}) implies that in the vacuum case $m_{\rm L}$ is constant $m_{\rm L}\equiv M$.
Here we show the Jebsen-Birkhoff theorem claiming that $C^2$ vacuum solutions with the symmetry under consideration are classified into; (i) Schwarzschild-Tangherlini-type solution; (ii) Nariai-type solution; (iii) special degenerate vacuum solution; and (iv) exceptional vacuum solution, where the third solution does not admit any hypersurface-orthogonal Killing vector.
In the exceptional vacuum solution, the metric on $(M^2, g_{ab})$ is totally {\it arbitrary}, which is realized in the case of $(Dr)^2=0$.
Indeed, the Jebsen-Birkhoff theorem in Lovelock gravity was proved in~\cite{zegers2005}. 
The original features of our proof are (I) we present the case with $(Dr)^2=0$ and (II) we clarify the relation between the realization of the special degenerate vacuum solution or exceptional vacuum solution and the degeneracy of vacua.
First we present all the vacuum solutions and then prove the theorem.

Note that if the assumption of twice differentiability of the metric is relaxed, the Jebsen-Birkhoff theorem is violated. 
Explicit examples were found in~\cite{Garraffo:2007fi} and \cite{Gravanis:2010zs}. 
In the latter, impulsive spherically symmetric gravitational shock waves were found.
Moreover, it is worth mentioning that the Jebsen-Birkhoff theorem is also valid for a certain class of higher derivative theories whose field equations reduce to second order when evaluated on the symmetric spacetimes considered here~\cite{Oliva:2011xu}.

\subsection{Maximally symmetric solution}
The maximally symmetric solution, namely Minkowski, dS or AdS solution, corresponds to $M=0$.
The maximally symmetric solution is given by 
\begin{align}
ds^2=&-\biggl(k-{\tilde \lambda}r^2\biggl)dt^2+\frac{dr^2}{k-{\tilde \lambda}r^2}+r^2\gamma_{ij}dz^i dz^j,\label{vacuum} 
\end{align}
where ${\tilde \lambda}:=2\lambda/[(n-1)(n-2)]$ and $\lambda$ is the effective cosmological constant.
Equation (\ref{qlm}) with $m_{\rm L}\equiv M=0$ gives an algebraic equation for ${\tilde \lambda}$;
\begin{align}
\label{lambda}
0=\sum_{p=0}^{[n/2]}{\tilde \alpha}_{(p)}{\tilde \lambda}^p=:v({\tilde\lambda}).
\end{align}  
This equation shows that the Minkowski vacuum ($\lambda=0$) is possible only for ${\alpha}_{(0)}=0$.

\begin{lm}
\label{lm:Min}
The Minkowski vacuum exists if and only if ${\alpha}_{(0)}=0$.
\end{lm}


If Eq.~(\ref{lambda}) does not allow any real ${\tilde \lambda}$, there is no maximally symmetric solution.
Proposition~\ref{lm:odd2} (and Corollary~\ref{coro:lovelock}) provides a sufficient condition for the existence of real ${\tilde \lambda}$.
Since Eq.~(\ref{lambda}) is a higher-order algebraic equation, there may be multiple real values of $\lambda$, each corresponding to a particular vacuum.
Let us consider when the degenerate vacua are realized.
We obtain
\begin{align}
\frac{dv}{d{\tilde\lambda}}=&\sum_{p=1}^{[n/2]}p{\tilde \alpha}_{(p)}{\tilde \lambda}^{p-1},\label{deg-vac}\\
\frac{d^2v}{d{\tilde\lambda}^2}=&\sum_{p=2}^{[n/2]}p(p-1){\tilde \alpha}_{(p)}{\tilde \lambda}^{p-2}.\label{deg-vac2}
\end{align}  
Hence, a {\it simply} generate vacuum is characterized by 
\begin{align}
\frac{dv}{d{\tilde\lambda}}=0,\label{deg-vac}
\end{align}  
while a {\it doubly} degenerate vacuum is characterized by
\begin{align}
\frac{dv}{d{\tilde\lambda}}=0,\qquad \frac{d^2v}{d{\tilde\lambda}^2}=0.
\end{align}  
In a similar manner, we can define a $q$th-order degenerate vacuum, where $q=1$ and $q=2$ correspond to the simply and doubly vacuum, respectively.
\begin{dn}
The $q$th-order degenerate vacuum is the maximally symmetric vacuum satisfying $d^sv/d{\tilde\lambda}^s=0$ for $s=1,2,\cdots,q$.
\end{dn}

In Lovelock gravity, $q \le [(n-3)/2]$ because ${\tilde \alpha}_{(n/2)}\equiv 0$ for even $n$.

\subsection{Schwarzschild-Tangherlini-type solution}
There also exists the following Schwarzschild-Tangherlini-type vacuum solution in Lovelock gravity;
\begin{align}
ds^2=-f(r)dt^2+\frac{dr^2}{f(r)}+r^2\gamma_{ij}dz^i dz^j. \label{f-vacuum}
\end{align}
Equation~(\ref{qlm}) gives the following master equation to determine the metric function $f(r)$~\cite{wheeler-lovelock};
\begin{align}
\label{alg}
{\tilde M} =\sum_{p=0}^{[n/2]}{\tilde \alpha}_{(p)}r^{n-1-2p}[k-f(r)]^p,
\end{align}
where ${\tilde M}:=2\kappa_n^2M/[(n-2)V_{n-2}^{(k)}]$.
For $M=0$, the solution (\ref{f-vacuum}) is a maximally symmetric solution given by (\ref{vacuum}).

\subsection{Special degenerate vacuum solution}
There are two special classes of vacuum solutions which do not have any hypersurface-orthogonal Killing vector.
The first class was originally found in quadratic Lovelock gravity, namely in Einstein-Gauss-Bonnet gravity, and given the name ``type-I vacuum solution''~\cite{cd2002}.
In the present paper, we call this solution the ``special degenerate vacuum solution'', for the following reason.
The metric of this solution is given by 
\begin{align}
ds^2=&-(k-\sigma r^2)e^{2\delta(t,r)}dt^2+\frac{dr^2}{k-\sigma r^2}+r^2\gamma_{ij}dz^i dz^j, \label{type-I} 
\end{align}  
where $\delta(t,r)$ is an {\it arbitrary} function.
The constant $\sigma$ is given by $\sigma={\tilde\lambda}$ satisfying Eqs.~(\ref{lambda}) and (\ref{deg-vac}).
This special degenerate vacuum solution is realized if the Lovelock coupling constants are tuned to admit degenerate vacua.
For $\delta(t,r)\equiv 0$, this solution reduces to the maximally symmetric solution with degenerate vacua.

Here we emphasize that the vacuum is not necessarily of fully degenerate vacuum, namely the $[(n-3)/2]$-th order degenerate vacuum.
This special degenerate vacuum solution is realized even when the Lovelock coupling constants admit a simply degenerate vacuum.

\subsection{Exceptional vacuum solution}
In this and the next subsections, we study the second class of vacuum solutions which do not have any hypersurface-orthogonal Killing vector is realized in the case of $(Dr)^2=0$, which is realized if $D^a r$ is a null vector or zero vector (namely $r$ is constant).
First we consider the case when $D^ar$ is a null vector and show that there is an exceptional class of vacuum solutions for $k=\alpha_{0}=\alpha_1=0$, in which the metric on $(M^2, g_{ab})$ is totally arbitrary.

It is shown by direct calculations that $D^2r=0$ if $D^ar$ is a null vector.
Also, $(D^a D_b r)(D^bD_a r)=0$ as shown below.
Using the following equation
\begin{align}
0=& D^aD_a((Dr)^2), \nonumber \\
=& 2 (D^aD_aD^b r)D_br+2 (D_aD^br)(D^aD_br),\nonumber \\
=& 2 (D_aD_bD^a r)D^br+2 (D_aD^br)(D^aD_br),
\end{align}
we show
\begin{align}
(D_aD^br)(D^aD_b r) =& -(D_aD_bD^a r)D^br, \nonumber \\
 =& - \frac12 {}^{(2)}R D_b rD^b r, \nonumber \\
=&0,
\end{align}
where we used the contracton of $D_aD_bD^d r-D_bD_aD^dr={}^{(2)}R^d_{~fab}D^fr$ together with the two-dimensional identity ${}^{(2)}R_{abcd}={}^{(2)}R(g_{ac}g_{bd}-g_{ad}g_{bc})$.

As a result, the field equations reduce to
\begin{align}
\kappa_n^2 T_{ab}=&\sum_{p=0}^{[n/2]}(n-2){\tilde \alpha}_{(p)}\biggl(-p\frac{D_aD_br}{r}\left(  \frac{k}{r^2} \right)^{p-1}- \frac{(n-1-2p)}{2}g_{ab}\left(  \frac{k}{r^2} \right)^{p}\biggl),\label{beq-ex3} \\
\kappa_n^2 T^{i}_{~~j}=&-\sum_{p=0}^{[n/2]}\frac{p{\tilde\alpha}_{(p)}}{2}\biggl(\frac{(n-2p-1)(n-2p-2)}{p} \left(  \frac{k}{r^2} \right)^{p}+ \overset{(2)}{R}{}\left(  \frac{k}{r^2} \right)^{p-1}\biggl)\delta^i_{~~j}.\label{beq-ex4}
\end{align}
For $k=0$, they are further simplified as
\begin{align}
\kappa_n^2 T_{ab}=&-(n-2)\biggl({\tilde \alpha}_{(1)}\frac{D_aD_br}{r}+ {\tilde \alpha}_{(0)}\frac{(n-1)}{2}g_{ab}\biggl), \label{beq-ex1}\\
\kappa_n^2 T^{i}_{~~j}=&-\biggl(\frac{(n-1)(n-2){\tilde\alpha}_{(0)}}{2}+\frac{{\tilde\alpha}_{(1)}}{2} \overset{(2)}{R}\biggl)\delta^i_{~~j}.\label{beq-ex2}
\end{align}
In the vacuum case, the trace of Eq.~(\ref{beq-ex1}) gives $\alpha_{0}=0$ (which implies that $m_{\rm L}$ is identically zero) and the field equations become
\begin{align}
{\alpha}_{(1)}D_aD_br=0,\qquad  {\alpha}_{(1)}\overset{(2)}{R}=0.\label{ex-last}
\end{align}
If $k=\alpha_{(0)}=0$ and $\alpha_{1}\ne 0$, we obtain $D_aD_br=0={}^{(2)}R$, which implies $R^\mu_{~\nu\rho\sigma}=0$ by the expressions (\ref{eq:Riemann}), and hence the spacetime is Minkowski.
On the other hand, if $k=\alpha_{0}=\alpha_{1}=0$, $(M^2, g_{ab})$ is totally arbitrary.
This is the exceptional vacuum solution.

\subsection{Nariai-type solution}
The last class of vacuum solutions is the Nariai-type solution, which is a cross product of two maximally symmetric manifolds with a constant warp factor, namely $(M^2, g_{ab})$ is maximally symmetric with $r=r_0$, where $r_0$ is a constant.
${}^{(2)}R$ and $r_0^2$ satisfy the following two algebraic equations:
\begin{align}
0=&\sum_{p=0}^{[n/2]}\frac{(n-2p-1){\tilde\alpha}_{(p)}k^p}{r_0^{2p}},\label{nari1}\\
\overset{(2)}{R}=&-\frac{\sum_{p=0}^{[n/2]}(n-2p-1)(n-2p-2){\tilde\alpha}_{(p)}(k/r_0^2)^{p}}{\sum_{q=0}^{[n/2]}q{\tilde\alpha}_{(q)}(k/r_0^2)^{q-1}}.\label{nari2}
\end{align}
Since ${}^{(2)}R$ is constant, $(M^2, g_{ab})$ is maximally symmetric.
This is the Nariai-type solution and the parameters must be chosen in such a way that $r_0^2$ is real and positive for a physically meaningful solution.

The Nariai-type solution contains Minkowski only if $k=0$.
In the case of $k=0$, Eqs.~(\ref{nari1}) and (\ref{nari2}) give
\begin{align}
{\alpha}_{(0)}=0, \qquad {\alpha}_{(1)}\overset{(2)}{R}=0.
\end{align}
If ${\alpha}_{(1)}\ne 0$, we obtain ${}^{(2)}R=0$ and the spacetime is Minkowski.
If ${\alpha}_{(1)}=0$, $(M^2, g_{ab})$ is totally arbitrary, which is the exceptional vacuum solution.

Actually, it is also possible for $k\ne 0$ that the denominator in Eq.~(\ref{nari2}) is vanishing and $(M^2, g_{ab})$ is arbitrary.
This is also the exceptional vacuum solution.
For $k\ne 0$, it is realized if 
\begin{align}
\sum_{p=0}^{[n/2]}{\tilde\alpha}_{(p)}x^p=\sum_{p=0}^{[n/2]}p{\tilde\alpha}_{(p)}x^p=\sum_{p=0}^{[n/2]}p^2{\tilde\alpha}_{(p)}x^p=0, \label{condition-ex}
\end{align}
where $x:=k/r_0^{2}$.
Using the function $v(x)$ defined by Eq.~(\ref{lambda}), we rewrite the above condition as
\begin{align}
v(x)=\frac{dv(x)}{dx}=\frac{d^2v(x)}{dx^2}=0.
\end{align}
In other words, if the theory admits a doubly degenerate vacuum, the exceptional vacuum solution with $k\ne 0$ is possible.
Since $k/r_0^{2}$ is related to the effective cosmological constant ${\tilde\lambda}$ of the doubly degenerate vacuum as ${\tilde\lambda}=k/r_0^{2}$, the following proposition is proved.
\begin{Prop}
({\it Exceptional vacuum solution.}) 
\label{prop:exceptional}
In Lovelock gravity admitting a doubly degenerate vacuum with a positive (negative) cosmological constant, there is the exceptional vacuum solution with constant $r$ for $k=1$ ($k=-1$).
In Lovelock gravity with ${\alpha}_{(0)}={\alpha}_{(1)}=0$, there is the exceptional vacuum solution with $(Dr)^2=0$ for $k=0$.
\end{Prop}

\bigskip

\subsection{Proof of the Jebsen-Birkhoff theorem}
Now let us prove the following Jebsen-Birkhoff theorem.
\begin{Prop}
\label{th:JB}
({\it Jebsen-Birkhoff theorem.}) 
The $C^2$ vacuum spacetime represented by the metric (\ref{eq:ansatz}) in Lovelock gravity with $(Dr)^2\ne 0$ is given by the Schwarzschild-Tangherlini-type solution for the coupling constants not giving a degenerate vacuum.
For the coupling constants giving degenerate vacua, the special degenerate vacuum solution is also possible.
If $(Dr)^2=0$ with $k=0$, there is no solution for $\alpha_{(0)} \ne 0$, while the solution with $\alpha_{(0)}=0$ is either Minkowski or the exceptional vacuum solution for $\alpha_{(1)}\ne 0$ or $\alpha_{(1)}=0$, respectively.
If $(Dr)^2=0$ with $k\ne 0$, the solution is the Nariai-type solution for the coupling constants not giving a doubly degenerate vacuum.
For the coupling constants giving a doubly degenerate vacuum with a positive (negative) cosmological constant, the exceptional vacuum solution is possible for $k=1$ ($k=-1$).
\end{Prop}
\noindent
{\it Proof}.
In the vacuum case, the unified first law (\ref{1stlaw1}) gives $m_{\rm L}=M$, where $M$ is a constant.
We consider the mixed component of Eq.~(\ref{beq-simp}) is zero.

First we consider the case with $(Dr)^2\ne 0$.
In this case, we may adopt the following coordinates without loss of generality;
\begin{align}
ds^2=-h(t,r)e^{2\delta(t,r)}dt^2+\frac{dr^2}{h(t,r)}+r^2\gamma_{ij}dz^i dz^j.
\end{align}
Then, the $(t,r)$ or $(r,t)$ component gives
\begin{align}
-\frac{\partial h}{\partial t}\sum_{p=0}^{[n/2]}{\tilde \alpha}_{(p)}p\left(  \frac{k-h }{r^2} \right)^{p-1}=0, \label{key1}
\end{align}
while the $(t,t)$ or $(r,r)$ component gives
\begin{align}
\frac{\partial \delta}{\partial r}\sum_{p=0}^{[n/2]}{\tilde \alpha}_{(p)}p\left(  \frac{k-h }{r^2} \right)^{p-1}=0.\label{key2}
\end{align}  
The above two equations are satisfied if $h=g(r)$, where $g(r)$ is the solution of the following algebraic equation;
\begin{align}
\sum_{p=0}^{[n/2]}{\tilde \alpha}_{(p)}p\left(  \frac{k-g(r) }{r^2} \right)^{p-1}=0.\label{alg2}
\end{align}  
The solution of the above algebraic equation is given by $g(r)=k-\sigma r^2$, where $\sigma$ is a constant satisfying
\begin{align}
\sum_{p=0}^{[n/2]}{\tilde \alpha}_{(p)}p\sigma^{p-1}=0.\label{jb-1}
\end{align}  
Comparing the above equation with Eq.~(\ref{deg-vac}), we find that $\sigma$ is the value of the effective cosmological constant ${\tilde\lambda}$ for a degenerate vacuum.
The $(i,j)$ component of the field equation (\ref{besij}) with $h(t,r)=k-\sigma r^2$ gives
\begin{align}
{\cal G}^{i}_{~~j}=&-\frac{1}{2r}\biggl(\sum_{p=0}^{[n/2]}{\tilde\alpha}_{(p)}\sigma^{p-1}\biggl)H\delta^i_{~~j}=0,
\end{align}
where 
\begin{align}
H:=&(n-1)(n-2)\sigma r+2p\biggl[-r(k-\sigma r^2)\biggl\{\frac{d^2\delta}{dr^2}+\biggl(\frac{d\delta}{dr}\biggl)^2\biggl\}+\{3k-n(k-\sigma r^2)\}\frac{d\delta}{dr}\biggl].
\end{align}
Hence, it is concluded that the metric of the special degenerate vacuum solution (\ref{type-I}) with a constant $\sigma$ satisfying Eqs.~(\ref{jb-1}) and 
\begin{align}
\sum_{p=0}^{[n/2]}{\tilde\alpha}_{(p)}\sigma^{p}=0 \label{jb-2}
\end{align}
solves the vacuum Lovelock equations for arbitrary $\delta(t,r)$.
If $\delta(t,r)\equiv 0$, the spacetime is maximally symmetric (\ref{vacuum}) with an effective cosmological constant ${\tilde\lambda}=\sigma$.
Since the conditions (\ref{jb-1}) and (\ref{jb-2}) mean that the vacuum is (at least simply) degenerate, this special degenerate vacuum solution is realized only for the Lovelock coupling constants giving degenerate vacua.

In the case where $h$ does not satisfy Eq.~(\ref{alg2}), Eqs.~(\ref{key1}) and (\ref{key2}) give $h={\bar h}(r)$ and $\delta={\bar \delta}(t)$.
We can set ${\bar \delta}\equiv 0$ without loss of generality by rescaling the coordinate $t$.
Then, the function ${\bar h}$ is given by ${\bar h}=f(r)$ satisfying Eq.~(\ref{alg}) and the $(t,t)$ or $(r,r)$ component and the $(i,i)$ component give dependent equations.
This is the Schwarzschild-Tangherlini-type solution.
It is noted that this Schwarzschild-Tangherlini-type solution is also possible for the Lovelock coupling constants giving degenerate vacua.

Next we consider the case with $(Dr)^2=0$.
For $k=0$, the trace of Eq.~(\ref{beq-ex1}) gives $\alpha_{0}=0$ and the field equations reduce to Eq.~(\ref{ex-last}).
If $\alpha_{(1)}=0$, Lovelock equations are fulfilled for any metric on $(M^2, g_{ab})$, which is the exceptional vacuum solution.
If $\alpha_{(1)} \ne 0$, the spacetime is Minkowski.
For $k \ne 0$, $r$ must be constant because the trace of Eq.~(\ref{beq-ex3}) gives a contradiction if $D_a r$ is a null vector.
Then, the vacuum equations reduce to Eqs.~(\ref{nari1}) and (\ref{nari2}).
This is the Nariai-type solution.
For the coupling constants giving a doubly degenerate vacuum with a positive (negative) cosmological constant, the exceptional vacuum solution with constant $r$ is also possible for $k=1$ ($k=-1$).

\qed

\bigskip

The above Jebsen-Birkhoff theorem allows us to understand the existence of a variety of vacuum configurations in Lovelock gravity admitting degenerate vacua. 
On of the interesting configurations is the vacuum wormhole.
By the coordinate transformation $d{\bar\rho}=dr/\sqrt{k-\sigma r^2}$ and ${\bar \rho}=\rho+\pi/(2\sqrt{\sigma})$, the special degenerate vacuum metric (\ref{type-I}) is transformed into
\begin{align}
ds^2=-dt^2+d\rho^2+\frac{k}{\sigma}\cosh^2(\sqrt{-\sigma}\rho)\gamma_{ij}dz^i dz^j, \label{type-I-2} 
\end{align}  
where we fix the arbitrary function $\delta(t,r(\rho))$ such that $g_{tt}=-1$ for simplicity.
The above metric represents a wormhole for negative $\sigma$ and $k=-1$.
The degeneracy of the metric function is indeed solved when we consider a non-maximally symmetric submanifold~\cite{wormhole-L,dot2010}.

Another interesting configuration is the Lifshitz vacuum solution in Lovelock gravity, which has been recently discussed in the third-order theory~\cite{Lifshitz-L}.
We consider the following metric:
\begin{align}
ds^2=-\frac{r^{2Z}}{l^{2Z}}f(r)dt^2+\frac{l^2}{r^2}\frac{dr^2}{g(r)}+r^2\gamma _{ij}(z)\D z^i\D z^j,
\label{lifshitz}
\end{align} 
where $Z$ and $l$ are constant.
The spacetime is called the Lifshitz spacetime for $f(r)=g(r)=1$. 
Since we have $(Dr)^2=r^2g(r)/l^2$, Eq.~(\ref{qlm}) gives
\begin{align}
\label{lifshitz-condition}
M=\sum_{p=0}^{[n/2]}{\tilde \alpha}_{(p)}r^{n-1-2p}(k-l^{-2}r^2g(r))^p,
\end{align}  
where $M$ is constant.
The Lovelock equations are actually solved for $M=0$ and 
\begin{align}
\label{lifshitz-solution}
g(r)=1+\frac{kl^2}{r^2}
\end{align}  
with arbitrary $f(r)$, where $l$ satisfies 
\begin{align}
\label{lifshitz-condition2}
0=\sum_{p=0}^{[n/2]}{\tilde \alpha}_{(p)}(-l^{-2})^p.
\end{align}  
Since $f(r)$ is arbitrary, this solution belongs to the special degenerate vacuum solution.
Indeed, the solution can be written as
\begin{align}
ds^2=-\frac{r^{2Z}}{l^{2Z}}f(r)dt^2+\frac{dr^2}{k+r^2/l^2}+r^2\gamma _{ij}(z)\D z^i\D z^j,
\end{align} 
which has the form of (\ref{type-I}) with $\sigma=-l^{-2}$ and appropriate $\delta(t,r)$.
For $f(r)=g(r)=1+kl^2/r^2$ with $k=-1$ and $l^2>0$, the spacetime represents an asymptotically Lifshitz topological vacuum black hole.
In~\cite{Lifshitz-L}, it was shown in the third-order Lovelock theory that there is the vacuum Lifshitz solution for a massive vector field, namely a Proca field, for the Lovelock coupling constants not giving degenerate vacua.

To close this section, we put one more comment on the relation between the ADM mass and the vacuum spacetime.
We proved the positive mass theorem in Section~\ref{subsec:qlm} (Proposition~\ref{coro:positivemass}). 
The ADM mass is zero in Minkowski spacetime.
Actually, under certain conditions, we can also show that the vanishing ADM mass means the flat spacetime, which is one of the consequences of the above Jebsen-Birkhoff theorem.
The following lemma is used in the proof.
(In the proof of the lemma, the dominant energy condition is used at the final step.)

\begin{lm}
\label{lm:Min-3}
Under the dominant energy condition, the asymptotically flat and $C^2$ solution with $m_{\rm L}\equiv 0$, $\alpha_{(1)}> 0$, and $\alpha_{(p)} \ge 0~(p \ne 1)$ is Minkowski.
\end{lm}
%
\noindent
{\it Proof}.
By using the expressions of the Riemann tensor (\ref{eq:Riemann}), we see that the solution with $(Dr)^2=0$ and $k\ne 0$ does not contain the Minkowski spacetime.
This is also true for $k=0$ in the case with $\alpha_{(1)}\ne 0$ by Proposition~\ref{th:JB}.
The special degenerate vacuum solution is not asymptotically flat unless $\alpha_{(1)}=0$ because Eq.~(\ref{deg-vac}) is not satisfied for ${\tilde\lambda}=0$.
Hence, for $\alpha_{(1)}\ne 0$, only the Schwarzschild-Tangherlini-type solution can be asymptotically flat among all the vacuum solutions.
Therefore, the spherically symmetric, asymptotically flat, and vacuum $C^2$ solution with $m_{\rm L}\equiv 0$ and $\alpha_{(1)}\ne 0$ is Minkowski.
Then, the lemma follows from Proposition~\ref{lm:zeromass}.

\qed

\begin{Prop}
\label{coro:positivemass2}
({\it Equivalence between zero ADM mass and Minkowski.}) 
Suppose the dominant energy condition and $\alpha_{(1)}>0$ and $\alpha_{(p)} \ge 0~(p\ne 1)$ are satisfied in the spherically symmetric, asymptotically flat, $C^2$ differentiable, and regular spacetime.
Then, the ADM mass is vanishing if and only if the spacetime is Minkowski.
\end{Prop}
\noindent
{\it Proof}.
Since the ADM mass is zero in the Minkowski spacetime, it is sufficient to show that the spacetime is Minkowski if the ADM mass is zero.
Because the quasi-local mass on a marginal surface is positive for $\alpha_{(1)}>0$ and $\alpha_{(p)} \ge 0~(p\ne 1)$ with $k=1$ by Eq.~(\ref{qlm-horizon}), there is no marginal surface on a spacelike hypersurface for vanishing ADM mass by Propositions~\ref{th:asymptotics} and \ref{th:monotonicity} and the fact that the asymptotically flat region consists of untrapped surfaces.
Then, by the same argument in the proof of Proposition~\ref{coro:positivemass}, the spacetime consists of untrapped surfaces and there is a regular center.
Hence, by Propositions~\ref{th:monotonicity} and \ref{th:positivity}, $m_{\rm L}\equiv 0$ is satisfied in the whole spacetime and finally the spacetime is Minkowski by Lemma~\ref{lm:Min-3}.

\qed


\section{Schwarzschild-Tangherlini-type black holes}
\label{sec6}
In this section, we discuss the properties of the Schwarzschild-Tangherlini-type vacuum black holes described by Eq.~(\ref{f-vacuum}) in more detail.

\subsection{Asymptotics}
Let us first consider the asymptotic behavior for $r\to \infty$.
By Lemma~\ref{lm:Min}, the asymptotically (locally) flat case is realized if and only if $\alpha_{(0)}=0$.
Then, from Eq.~(\ref{alg}), we obtain the behavior of the metric function around $r\to \infty$ as
\begin{align}
f(r)\simeq k-\frac{{\tilde M}}{{\tilde \alpha}_{(1)}r^{n-3}}.
\end{align}
Hence, the asymptotic flatness condition is satisfied, which is consistent with Proposition~\ref{th:asymptotics}.

Next we consider the asymptotically (A)dS case.
Using Eqs.~(\ref{alg}) and (\ref{lambda}), we obtain
\begin{align}
f(r)\simeq k-\frac{{\tilde M}}{\sum_{p=0}^{[n/2]}{\tilde \alpha}_{(p)}p{\tilde\lambda}^{p-1}r^{n-3}}-{\tilde\lambda}r^2,\label{fallAdS}
\end{align}
where ${\tilde\lambda}(\ne 0)$ is defined in Eq.~(\ref{vacuum}).
Indeed, the denominator of the mass term may be zero in some case.
In such a case, the fall-off rate to $r\to \infty$ is slower than the above.
Comparing with Eq.~(\ref{deg-vac}), we find that it occurs for degenerate vacua.
\begin{Prop}
\label{th:asmpAdS}
({\it Slow fall-off to (A)dS infinity.}) 
In the asymptotically (locally) degenerate (A)dS Schwarzschild-Tangherlini-type spacetime (\ref{f-vacuum}), the fall-off rate to $r\to \infty$ is slower than in Eq.~(\ref{fallAdS}).
\end{Prop}

\subsection{Singularities}
Next we consider the existence of curvature singularities.
The Kretschmann invariant is given by
\begin{align}
K:=&R_{\mu\nu\rho\sigma}R^{\mu\nu\rho\sigma}, \nonumber \\
=&\biggl(\frac{d^2f}{dr^2}\biggl)^2+\frac{2(n-2)}{r^2}\biggl(\frac{df}{dr}\biggl)^2+\frac{2(n-2)(n-3)}{r^4}(k-f)^2.
\end{align}
Because the coefficients of all the terms are positive definite, $K$ blows up if either $f$, $df/dr$, or $d^2f/dr^2$ blows up.
In higher-order Lovelock gravity, there are two types curvature singularities.
One is a central singularity at $r=0$ and the other is a so-called branch singularity located at $r=r_{\rm b}>0$.
Unlike in the general relativistic case, the Schwarzschild-Tangherlini-type spacetime (\ref{f-vacuum}) in Lovelock gravity admits such a non-central curvature singularity.
Differentiating Eq.~(\ref{alg}), we obtain 
\begin{align}
\frac{df}{dr}=\frac{\sum_{p=0}^{[n/2]}{\tilde \alpha}_{(p)}(n-1-2p)r^{n-2-2p}[k-f(r)]^p}{\sum_{q=0}^{[n/2]}{\tilde \alpha}_{(q)}qr^{n-1-2q}[k-f(r)]^{q-1}}.\label{dfdr}
\end{align}
Hence, the position of the branch singularity is determined by 
\begin{align}
0=\sum_{q=0}^{[n/2]}{\tilde \alpha}_{(q)}qr_{\rm b}^{-2q}[k-f(r_{\rm b})]^{q-1} \label{branch}
\end{align}
with non-zero numerator in the right-hand side of Eq.~(\ref{dfdr}).
It is easy to show that the spacetime is maximally symmetric if both the denominator and numerator are vanishing.
This branch singularity appears even in the quadratic Einstein-Gauss-Bonnet gravity~\cite{tm2005}, where $f(r_{\rm b})$ is finite but $df/dr(r_{\rm b})$ blows up.

If there is a branch singularity, it is quite a non-trivial task to identify the parameter region for black holes in Lovelock gravity.
This is because, even though there is an outer Killing horizon, it is not an event horizon if there is a branch singularity outside.
Hence, an outermost outer Killing horizon at $r=r_{\rm h}$ is an event horizon only if the metric is real and there is no singularity for $r_{\rm h}<r<\infty$.

First let us study the central singularity.
One may think that the Lovelock higher-curvature effect cures the central singularity in the Schwarzschild-Tangherlini spacetime in general relativity.
However, it is shown that there exists a central curvature singularity for $M\ne 0$.
\begin{Prop}
\label{th:c-sing}
({\it Central curvature singularity.}) 
In the Schwarzschild-Tangherlini-type spacetime (\ref{f-vacuum}), there is a curvature singularity at $r=0$ if $M \ne 0$.
This central singularity becomes null only in odd dimensions with ${\tilde \alpha}_{((n-1)/2)}\ne 0$ and ${\tilde M}=k^{(n-1)/2}{\tilde \alpha}_{((n-1)/2)}$.
\end{Prop}
\noindent
{\it Proof}.
The asymptotic behavior of the metric function around $r=0$ is given from Eq.~(\ref{alg}) as
\begin{align}
f(r)\simeq k-\biggl(\frac{{\tilde M}}{{\tilde \alpha}_{(s)}r^{n-1-2s}}\biggl)^{1/s},\label{r=0oddeven}
\end{align}
where $s$ is the largest integer with non-zero ${\tilde \alpha}_{(s)}$ in $1 \le s \le [n/2]$.
In odd dimensions, $f(r)$, and hence $K$, blows up around $r=0$ for $s \ne (n-1)/2$.
For $s =(n-1)/2$, $f(r)$ is finite there but different from $k$ and hence $K$ blows up.
In even dimensions, $K$ always blows up because of ${\tilde \alpha}_{(n/2)}\equiv 0$.

If $f(r)$ behaves $f(r)\simeq f_0r^d$ around $r=0$, the central singularity is non-null and null for $d<1$ and $d\ge 1$, respectively.
Therefore, the central singularity is non-null unless $s =(n-1)/2$ with $k=({\tilde M}/{\tilde \alpha}_{((n-1)/2)})^{2/(n-1)} \ne 0$.
For $s =(n-1)/2$ with $k=({\tilde M}/{\tilde \alpha}_{((n-1)/2)})^{2/(n-1)}$, $f(r)$ behaves around $r=0$ as
\begin{align}
f(r)\simeq f_0r^d
\end{align}
with $d\ge 2$~\footnote{If ${\tilde \alpha}_{((n-3)/2)}\ne 0$, we have $d=2$ and $f_0 =2{\tilde \alpha}_{((n-3)/2)}/[(n-1){\tilde \alpha}_{((n-1)/2)}]$.}.
Hence, the central singularity is null in this case.
\qed

\bigskip

From the expression~(\ref{r=0oddeven}), we obtain the following condition for the region around $r=0$ to be contained in an unphysical region.
\begin{Prop}
\label{th:r=0}
({\it Unphysical central region.}) 
Let $s$ be the largest integer with non-zero ${\tilde \alpha}_{(s)}$ in $1 \le s \le [n/2]$.
In the Schwarzschild-Tangherlini-type spacetime (\ref{f-vacuum}), if $s$ is even and ${\tilde M}/{\tilde \alpha}_{(s)}<0$, the metric is complex around $r=0$.
\end{Prop}

\bigskip

In general Lovelock gravity, where $s=(n-1)/2$ and $s=(n-2)/2$ in odd and even dimensions, respectively, the metric is complex around $r=0$ for $n=4q+1$ or $n=4q+2$ if ${\tilde M}/{\tilde \alpha}_{(s)}<0$.
The above proposition claims that for positive $M$ and the Lovelock coupling constants, the metric is real around $r=0$.
In the presence of an electric charge, however, the situation is different. (See Appendix C.)

In the Schwarzschild-Tangherlini spacetime ($k=1$) with ${\alpha}_{(1)}>0$ (positive Newton constant), a central timelike naked singularity appears only for negative mass.
However, in Lovelock gravity, it is different in odd and even dimensions shown by Eq.~(\ref{r=0oddeven}).
\begin{Prop}
\label{th:positivemassNS}
({\it Positive mass naked singularity in odd dimensions.}) 
Let $s$ be the largest integer with non-zero ${\tilde \alpha}_{(s)}$ in $1 \le s \le [n/2]$ and suppose $\alpha_{(s)}>0$ in the Schwarzschild-Tangherlini-type spacetime (\ref{f-vacuum}) with $k=1$ and $M>0$.
Then, the central curvature singularity is spacelike if $s \ne (n-1)/2$.
If $s=(n-1)/2$, the central curvature singularity is timelike, null, and spacelike for $0<M<(n-2)V_{n-2}^{(1)}{\tilde\alpha}_{((n-1)/2)}/(2\kappa_n^2)$, $M=(n-2)V_{n-2}^{(1)}{\tilde\alpha}_{((n-1)/2)}/(2\kappa_n^2)$, and $M>(n-2)V_{n-2}^{(1)}{\tilde\alpha}_{((n-1)/2)}/(2\kappa_n^2)$, respectively.
\end{Prop}
 
\bigskip

The above proposition claims that positive mass singularities are in the trapped region in even dimensions as in the Schwarzschild-Tangherlini spacetime.
In contrast, in odd dimensions, locally naked singularities with positive mass are possible for ${\alpha}_{((n-1)/2)}>0$.
The upper bound of the mass of the naked singularity is given by $M=(n-2)V_{n-2}^{(1)}{\tilde\alpha}_{((n-1)/2)}/(2\kappa_n^2)$.
Here we note that the above proposition does not claim the existence of globally naked singularities in odd dimensions in Lovelock gravity. 
In order to show the global visibility of the singularity, it is necessary to show the existence of a single spacetime which extends over all values of $r$.
As explained in Section~\ref{sec:branch}, Proposition~\ref{lm:odd2} does not ensure that.
In order to answer the problem of global visibility of the singularity in the Schwarzschild-Tangherlini-type spacetime, a novel approach invented in~\cite{ce2011} must be useful.

Next we study the branch singularity.
The existence of the unphysical region with complex $r$ shown by Proposition~\ref{th:r=0} means that there is a positive lower bound for the value of $r$ in the coordinate system adopted.
In the quadratic case, namely in Einstein-Gauss-Bonnet gravity, there is a branch curvature singularity there~\cite{tm2005}. 
The following proposition claims that it is indeed the case in general Lovelock gravity.
\begin{Prop}
\label{th:branch-ex}
({\it Existence of branch singularities.}) 
If there is a positive lower or upper bound for the value of $r$ with real metric in the Schwarzschild-Tangherlini-type spacetime (\ref{f-vacuum}), then there is a branch singularity at the boundary.
\end{Prop}
\noindent
{\it Proof}.
For given values of ${\alpha}_{(p)}$, the master equation (\ref{alg}) can be identified as a polynomial function ${\tilde M}={\tilde M}(f)$ with coefficients depending on $r$.
Then the shape of the function ${\tilde M}(f)$ changes depending on the value of $r$.
The intersections with a given value of the parameter ${\tilde M}$ determines the values of $f$ at that areal radius $r$ in the Schwarzschild-Tangherlini-type solution.
If there are several intersections, it means that there are several branches of solutions at that value of $r$.
By assumption, there is a positive domain of $r$ admitting intersections.
If one decreases or increases $r$ continuously from such a value of $r$, some branches of solutions may coincide and disappear.
Indeed, if the metric becomes complex for some value of $r$, the corresponding branch must disappear.
This vanishing point is characterized by 
\begin{align}
\frac{d{\tilde M}}{df} =-\sum_{p=0}^{[n/2]}p{\tilde \alpha}_{(p)}r^{n-1-2p}(k-f)^{p-1}=0.\label{dMdf}
\end{align}
Comparing with Eq.~(\ref{branch}), we see that it corresponds to a curvature singularity or the spacetime is maximally symmetric.
However, the latter case contradicts with the assumption because the metric in the maximally symmetric spacetime is real for $r\ge 0$.
\qed

\bigskip

\subsection{The first law of the black-hole thermodynamics}
\label{subsec:thermo}
In the Schwarzschild-Tangherlini-type spacetime (\ref{f-vacuum}), a Killing horizon $r=r_{\rm h}$ is given by $f(r_{\rm h})=0$.
An outer Killing horizon is defined by $df/dr|_{r=r_{\rm h}}>0$, while a degenerate Killing horizon is defined by $df/dr|_{r=r_{\rm h}}=0$.
A black-hole event horizon corresponds to the outermost outer Killing horizon if the metric is real and there is no singularity for $r_{\rm h}<r<\infty$.
A degenerate horizon satisfying $d^2f/dr^2|_{r=r_{\rm h}}>0$ may also be a black-hole event horizon.
As mentioned in the previous subsection, it is rather tedious to specify the parameter region admitting the black-hole configuration in Lovelock gravity because of the branch singularity.
In this paper, we just assume the existence of an outer Killing horizon in the physical domain of the coordinate $r$.

Here we show that the first law of the black-hole thermodynamics is satisfied.
Using Eq.~(\ref{alg}), we obtain the relation between the mass parameter and the horizon radius as
\begin{align}
\label{m-r}
M&=\frac{(n-2)V_{n-2}^{(k)}}{2\kappa_n^2}\sum_{p=0}^{[n/2]}{\tilde \alpha}_{(p)}r_{\rm h}^{n-1-2p}k^p.
\end{align}  
Variation of the above equation with respect to the variable $r_{\rm h}$ gives
\begin{align}
\delta M =&\frac{(n-2)V_{n-2}^{(k)}}{2\kappa_n^2}\sum_{p=0}^{[n/2]}(n-1-2p){\tilde \alpha}_{(p)}r_{\rm h}^{n-2-2p}k^p\delta r_{\rm h}.\label{id2}
\end{align} 
The surface gravity $\kappa$ on the Killing horizon is defined by $\kappa:=(1/2)(df/dr)|_{r=r_{\rm h}}$, which is given by evaluating Eq.~(\ref{dfdr}) on the horizon as
\begin{align}
\kappa=&\frac{\sum_{p=0}^{[n/2]}{\tilde \alpha}_{(p)}(n-1-2p)r_{\rm h}^{n-2-2p}k^p}{\sum_{q=0}^{[n/2]}{\tilde \alpha}_{(q)}2q r_{\rm h}^{n-1-2q}k^{q-1}}.
\end{align}  
Using the temperature of the horizon defined by $T:=\kappa/(2\pi)$, we rewrite Eq.~(\ref{id2}) as
\begin{align}
\delta M &= \frac{2\pi(n-2)V_{n-2}^{(k)}T}{\kappa_n^2}\sum_{p=0}^{[n/2]}{\tilde \alpha}_{(p)}p r_{\rm h}^{n-1-2p}k^{p-1}\delta r_{\rm h}.\label{deltaM}
\end{align}

On the other hand, the entropy on the Killing horizon is calculated using the Wald formula.
The Wald entropy is defined by the following integral performed on $(n-2)$-dimensional spacelike bifurcation surface $\Sigma$~\cite{wald1993,iyerwald1994,jkm1994}:
\begin{align}
S_{\rm W}:=&-2\pi \oint\left(\frac{\partial {\cal L}}{\partial R_{\mu\nu\rho\sigma}}\right)\varepsilon_{\mu\nu}\varepsilon_{\rho\sigma}dV_{n-2},
\end{align}
where $dV_{n-2}$ is the volume element on $\Sigma$, $\varepsilon_{\mu\nu}$ is the binormal vector to $\Sigma$ normalized as $\varepsilon_{\mu\nu}\varepsilon^{\mu\nu}=-2$, and ${\cal L}$ is the Lagrangian density.
$\varepsilon_{\mu\nu}$ may be written as $\varepsilon_{\mu\nu}=\xi_{\mu}\eta_{\nu}-\eta_{\mu}\xi_{\nu}$, where $\xi^\mu$ and $\eta^\mu$ are two null vector fields normal to $\Sigma$ satisfying $\xi_\mu\eta^\mu=1$.
In Lovelock gravity, ${\cal L}$ is given by
\begin{align}
{\cal L}=&\frac{1}{2\kappa_n^2}\sum_{p=0}^{[n/2]}\alpha_{(p)}{\ma L}_{(p)},\\
=&\frac{1}{2\kappa_n^2}\sum_{p=0}^{[n/2]}\frac{\alpha_{(p)}}{2^p}\delta^{\mu_1\cdots \mu_p\nu_1\cdots \nu_p}_{\rho_1\cdots \rho_p\sigma_1\cdots \sigma_p}R_{\mu_1\nu_1}^{\phantom{\mu_1}\phantom{\nu_1}\rho_1\sigma_1}\cdots R_{\mu_p\nu_p}^{\phantom{\mu_p}\phantom{\nu_p}\rho_p\sigma_p}.
\end{align}

For the metric (\ref{f-vacuum}), $\Sigma$ is given by $t=$constant and $r=r_{\rm h}=$constant and we have $\varepsilon_{tr}=1$.
The Wald entropy is calculated to give
\begin{align}
S_{\rm W}=&-\frac{2\pi}{16\pi G_n} \oint\left(\frac{\partial {\cal L}}{\partial R_{abcd}}\right)\varepsilon_{ab}\varepsilon_{cd}r^{n-2}d\Omega_{n-2}^{(k)},\nonumber \\
=&\frac{2\pi}{\kappa_n^2}\sum_{p=0}^{[n/2]}{\tilde \alpha}_{(p)}\frac{p(n-2)}{n-2p} k^{p-1}V_{n-2}^{(k)}r_{\rm h}^{n-2p}, \label{waldS}
\end{align}
where $d\Omega_{n-2}^{(k)}$ is the volume element on $(K^{n-2},\gamma_{ij})$.
This is the same form as the dynamical entropy (\ref{S}).
In the static case with $k=1$, the expression (\ref{waldS}) reduces to the black-hole entropy derived by Whitt~\cite{whitt1988}.

Varying the Wald entropy with respect to $r_{\rm h}$, we obtain
\begin{align}
\delta S_{\rm W}=&\frac{2\pi(n-2)V_{n-2}^{(k)}}{\kappa_n^2}\sum_{p=0}^{[n/2]}{\tilde \alpha}_{(p)}p k^{p-1}r_{\rm h}^{n-2p-1}\delta r_{\rm h}. \label{deltaS}
\end{align}  
Using Eqs.~(\ref{deltaM}) and (\ref{deltaS}), we finally obtain
\begin{align}
\delta M = T\delta S_{\rm W}.
\end{align}  
This is the first-law of the black-hole thermodynamics in Lovelock gravity.

The Wald entropy (\ref{waldS}) is a higher-order polynomial.
Hence, one may think that there may be extrema of $S_{\rm W}$.
The following proposition claims that branch singularities always exist there. 
\begin{Prop}
\label{th:c-sing}
({\it Branch singularity and extremum of the black-hole entropy.}) 
There is a branch singularity on the Killing horizon corresponding to an extremum of the black-hole entropy (\ref{waldS}) in the Schwarzschild-Tangherlini-type spacetime.
\end{Prop}
\noindent
{\it Proof}.
By the expression (\ref{deltaS}), the extremum of the black-hole entropy is given by 
\begin{align}
\sum_{p=0}^{[n/2]}{\tilde \alpha}_{(p)}p k^{p-1}r_{\rm h}^{n-2p-1}=0.
\end{align}  
From the expression of (\ref{dfdr}), it is shown that $df/dr$ blows up there, which corresponds to a branch singularity.

\qed


\subsection{Thermodynamical quantities}

In this subsection, we present the expressions of the thermodynamical quantities for Lovelock black holes represented by Eq.~(\ref{f-vacuum}).
The heat capacity $C$ is given by
\begin{align}
C:=\frac{dM}{dT}=\frac{dM}{dr_{\rm h}}\biggl/\frac{dT}{dr_{\rm h}},
\end{align}
where
\begin{align}
\frac{dT}{dr_{\rm h}}=&\frac{1}{2\pi}\biggl(\sum_{p=0}^{[n/2]}{\tilde \alpha}_{(p)}2p r_{\rm h}^{n-1-2p}k^{p-1}\biggl)^{-2} \nonumber \\
&\times\biggl[\biggl(\sum_{q=0}^{[n/2]}{\tilde \alpha}_{(q)}2q r_{\rm h}^{n-1-2q}k^{q-1}\biggl)\biggl\{\sum_{s=0}^{[n/2]}{\tilde \alpha}_{(s)}(n-1-2s)(n-2-2s)r_{\rm h}^{n-3-2s}k^s\biggl\} \nonumber \\
&-\biggl\{\sum_{q=0}^{[n/2]}{\tilde \alpha}_{(q)}2q(n-1-2q) r_{\rm h}^{n-2-2q}k^{q-1}\biggl\}\biggl\{\sum_{s=0}^{[n/2]}{\tilde \alpha}_{(s)}(n-1-2s)r_{\rm h}^{n-2-2s}k^s\biggl\}\biggl].
\end{align}
and
\begin{align}
\frac{dM}{dr_{\rm h}}=&\frac{(n-2)V_{n-2}^{(k)}}{2\kappa_n^2}\sum_{p=0}^{[n/2]}{\tilde \alpha}_{(p)}(n-1-2p)r_{\rm h}^{n-2-2p}k^p.
\end{align}  
Positive heat capacity implies that the black hole is locally thermodynamically stable. However, to analyze the global thermodynamical stability one must obtain the free energy of the black hole defined by $F:=M-TS$.

Plugging in the expressions for mass, temperature and entropy for the Schwarzschild-Tangherlini-type static black holes (\ref{f-vacuum}), we obtain
\begin{align}
F=&\frac{(n-2)V_{n-2}^{(k)}}{2\kappa_n^2}\biggl[\sum_{p=0}^{[n/2]}{\tilde \alpha}_{(p)}r_{\rm h}^{n-1-2p}k^p \nonumber \\
&-\frac{\{\sum_{p=0}^{[n/2]}{\tilde \alpha}_{(p)}(n-1-2p)r_{\rm h}^{n-2-2p}k^p\}\{\sum_{q=0}^{[n/2]}{\tilde \alpha}_{(q)}q k^{q-1}(n-2q)^{-1}r_{\rm h}^{n-2q}\}}{\sum_{s=0}^{[n/2]}{\tilde \alpha}_{(s)}s r_{\rm h}^{n-1-2s}k^{s-1}}\biggl].
\label{free}
\end{align}  
This gives an expression for the free energy in terms of the horizon radius and the coupling constants. In the simplest case of Schwarzschild-AdS black holes in four dimensions, Hawking and Page showed that beyond a critical temperature, a sufficiently large black hole has negative free energy with respect to the AdS background and thus is thermodynamically stable \cite{Hawking:1982dh}. This can be seen from (\ref{free}) which for Einstein's gravity with a cosmological constant reduces to
\begin{align}
 F=&\frac{V_{n-2}^{(1)}}{2\kappa_n^2}r_{\rm h}^{(n-3)}(1-{\tilde \alpha}_{(0)}r_{\rm h}^2),
\end{align}
where we set ${\alpha}_{(1)}=1$.
For ${\tilde \alpha}_{(0)}>0$, the free energy $F$ can be negative for a large black hole. 

The thermodynamical stability of a Lovelock black hole has been recently studied in~\cite{ce2011} using the free energy.
Indeed, for higher-order Lovelock gravity, the phase space is much more complicated and the expression (\ref{free}) for the free energy might not be very useful for analyzing the thermodynamical stability. 
In such a case, it is more convenient to group the different contributions to the free energy by using the Smarr law \cite{Kastor,Kastor:2009wy,Kastor:2010gq}.

\subsection{Toroidal black holes $k=0$}
Closing this section, we mention a bit more about the toroidal case with $k=0$ because in this case the formulae obtained above become drastically simple.
The mass-horizon relation (\ref{m-r}) becomes 
\begin{align}
M&=\frac{(n-2)V_{n-2}^{(0)}}{2\kappa_n^2}{\tilde \alpha}_{(0)}r_{\rm h}^{n-1},\label{m-rk=0}
\end{align}  
which is the same as the general relativistic case.
Since $M=0$ gives a maximally symmetric spacetime with $k=0$, ${\tilde \alpha}_{(0)}\ne 0$ is required for the existence of a Killing horizon.
Evaluating Eq.~(\ref{dfdr}) on the horizon $r=r_{\rm h}$ with $k=0$, we obtain
\begin{align}
{\tilde\alpha}_{(1)}\frac{df}{dr}\biggl|_{r=r_{\rm h}}=(n-1){\tilde\alpha}_{(0)}r_{\rm h}.
\end{align}  
There is no Killing horizon for ${\alpha}_{(1)}=0$ because the above equation gives a contradiction ${\alpha}_{(0)}=0$.
In summary, there is no Killing horizon for ${\alpha}_{(0)}{\alpha}_{(1)}=0$ and there is an outer Killing horizon for ${\alpha}_{(0)}{\alpha}_{(1)}>0$.

Hereafter we assume ${\alpha}_{(0)}{\alpha}_{(1)}\ne 0$.
The Wald entropy (\ref{waldS}) becomes
\begin{align}
S_{\rm W}=&\frac{2\pi {\tilde \alpha}_{(1)} V_{n-2}^{(0)}r_{\rm h}^{n-2}}{\kappa_n^2},
\end{align}
which is the same as the Bekenstein-Hawking formula.
The temperature $T$, heat capacity $C$, and the free energy $F$ become
\begin{align}
T=&\frac{(n-1){\tilde \alpha}_{(0)}r_{\rm h}}{4\pi{\tilde \alpha}_{(1)}},\\
C=&\frac{2\pi(n-2){\tilde \alpha}_{(1)}V_{n-2}^{(0)}r_{\rm h}^{n-2}}{\kappa_n^2},\\
F=&-\frac{{\tilde \alpha}_{(0)}V_{n-2}^{(0)}r_{\rm h}^{n-1}}{2\kappa_n^2}.
\end{align}  
Hence, for $\alpha_{(0)}>0$ and $\alpha_{(1)}> 0$, a toroidal black hole is thermodynamically stable.

\section{Summary}
\label{sec:summary}
In this paper, we have investigated some of the basic properties of symmetric spacetimes having the isometries of an $(n-2)$-dimensional maximally symmetric space in Lovelock gravity.
This is a generalization of the previous works in Einstein-Gauss-Bonnet gravity~\cite{mn2008,nm2008,maeda2008} and contains a large set of new results.

In section~\ref{sec3}, we have shown that the generalized Misner-Sharp mass proposed in~\cite{mn2008} satisfies the unified first law and is a natural generalization of the Misner-Sharp mass in general relativity. 
This quasi-local mass has been used to evaluate the mass of a dynamical black hole in the subsequent section.
In section~\ref{sec4}, we have shown that the properties of a dynamical black hole defined by a future outer trapping horizon in Lovelock gravity with ${\alpha}_{(p)}k^{p-1} \ge 0$ for $p\ge 1$ and $\sum_{p=1}^{[n/2]}{\tilde\alpha}_{(p)}k^{p-1} \ne 0$ are shared with the case in general relativity.  
In section~\ref{sec5}, we have proven the Jebsen-Birkhoff theorem that the vacuum solutions are classified into; (i) Schwarzschild-Tangherlini-type solution; (ii) Nariai-type solution; (iii) special degenerate vacuum solution; and (iv) exceptional vacuum solution.
It was shown that the special degenerate vacuum solution is realized only for the coupling constant giving degenerate vacua.
On the other hand, the exceptional vacuum solution is realized for $k=\alpha_{(0)}=\alpha_{(1)}=0$ or $k\ne 0$ with the coupling constants giving a doubly degenerate vacuum.

In section~\ref{sec6}, we have studied the Schwarzschild-Tangherlini-type black hole in more detail.
We have clarified the asymptotic behavior of the spacetime and the existence of curvature singularities. 
It was shown that its constant quasi-local mass satisfies the first law of black-hole thermodynamics using the Wald entropy.
We have derived the expressions for the heat capacity and free energy.
In the toroidal case ($k=0$), the formulae become drastically simple.
For the Schwarzschild-Tangherlini-type toroidal black hole, the higher-curvature effects do not appear in the expressions for the position of the horizon and the thermodynamical quantities, but introduce branch singularities in the spacetime.
It is interesting to see that, while most of the spacetime properties in Lovelock gravity are different in odd and even dimensions, several properties depend on the remainder of $n$ divided by four. 
(See Corollary~\ref{coro:lovelock} and Propositions~\ref{th:r=0} and \ref{th:DCmass}.)

The results presented in this paper could be a firm basis for the further investigations on Lovelock black holes.
However, we are dissatisfied with that we have clarified the properties of a dynamical black hole only for ${\alpha}_{(p)}k^{p-1} \ge 0$ for $p\ge 1$ and $\sum_{p=1}^{[n/2]}{\tilde\alpha}_{(p)}k^{p-1} \ne 0$ and leaving the other cases.
Indeed, in the quadratic case, namely Einstein-Gauss-Bonnet gravity, rigorous results have been obtained also in other cases by classifying the solutions into two branches depending on whether they have a general relativistic limit or not~\cite{nm2008,maeda2010}.
This classification is performed by solving the expression of the quasi-local mass (\ref{qlm}) for $(Dr)^2$.
However, in higher-order Lovelock gravity, it becomes a higher-order algebraic equation for $(Dr)^2$ and it is much more complicated to distinguish the branches of solutions.
In order to solve this problem, the new method invented in~\cite{ce2011} would be useful.

Another important subject for future investigations is to study less symmetric spacetimes.
A generalization of the present study is to consider a $(n-2)$-dimensional submanifold not being a maximally symmetric but some Einstein space.
In Einstein-Gauss-Bonnet gravity, the compatibility condition for the Einstein submanifold with the field equations was given by Dotti and Gleiser~\cite{dg2005}.
(See also~\cite{dot2010}.)
Intriguingly, as shown in~\cite{maeda2010}, the Misner-Sharp formalism can be generalized in such a case.
It strongly suggests the possibility of a further generalization in Lovelock gravity.
This is an interesting problem for future research.

\subsection*{Acknowledgements}
The authors thank Julio Oliva for fruitful discussions and comments. 
HM thanks Masato Nozawa, Jos\'e D. Edelstein, Tomohiro Harada, and Ricardo Troncoso for valuable comments.
HM would like also to thank the Cosmophysics group in KEK and the group of Gravitation and Cosmology at PUCV for hospitality and support.
This work has been partially funded by the Fondecyt grants 1100328, 1100755 (HM) and by the Conicyt grant "Southern Theoretical Physics Laboratory" ACT-91. 
This work was also partly supported by the JSPS Grant-in-Aid for Scientific Research (A) (22244030).
The Centro de Estudios Cient\'{\i}ficos (CECS) is funded by the Chilean Government through the Centers of Excellence Base Financing Program of Conicyt.

\appendix

\section{Tensor decomposition}
In this appendix, we present the decompositions of the curvature tensors.
The non-vanishing components of the Levi-Civit\'a connections are
\begin{align}
\begin{aligned}
{\Gamma ^a}_{bc}&=\overset{(2)}{\Gamma}{}^a_{bc }(y),\quad 
{\Gamma ^i}_{ij}={\hat{\Gamma} ^i}_{~jk}(z), \\
{\Gamma ^a}_{ij}&=-r (D^a r) \gamma _{ij},\quad 
{\Gamma ^i}_{ja}=\frac{D_a r}{r}{\delta ^i}_j, 
\end{aligned}
\end{align}
where the superscript (2) denotes the two-dimensional quantity,
and $D_a$ is the two-dimensional linear connection compatible with
$g_{ab}$. ${\hat \Gamma ^i}_{~jk}$ is the Levi-Civit\'a connection
associated with $\gamma _{ij}$.
The Riemann tensor is given by
\begin{align}
{R^a}_{bcd}&=\overset{(2)}{R}{}^a_{~~bcd},\nonumber \\
{R^a}_{ibj}&=-r(D^a D_b r)\gamma _{ij},
\label{eq:Riemann}\\
{R^i}_{jkl}&=[k-(Dr)^2]({\delta ^i}_k\gamma _{jl}
-{\delta ^i}_l\gamma _{jk}). \nonumber 
\end{align}
The Ricci tensor and the Ricci scalar are given by 
\begin{align}
R_{ab}&=\overset{(2)}{R}_{ab}-(n-2)\frac{D_aD_br}{r}, \nonumber \\
R_{ij}&=\left\{-r D^2r+(n-3)[k-(Dr)^2]\right\}\gamma _{ij}, \label{eq:Ricci} \\
R&=\overset{(2)}{R}-2(n-2)\frac{D^2r}{r}+(n-2)(n-3)\frac{k-(Dr)^2}{r^2}. \nonumber
\end{align}

\section{Lovelock gravity with a unique maximally symmetric vacuum}
\label{sec:dc}
The form of the quasi-local mass $m_{\rm L}$ becomes quite simple in the case where there is a unique maximally symmetric vacuum solution.
This is realized if ${\tilde \alpha}_{(p)}$ satisfies
\begin{eqnarray}
{\tilde \alpha}_{(p)}=\left\{
\begin{array}{ll}
\varepsilon^{-1}
\left(
\begin{array}{cc}
s \\
p
\end{array}
\right)l^{2p-2} \quad (\mbox{for}~p\le  s),\\
0~~~~~~~~~~~~~~~~~~~~~(\mbox{for}~p\ge s+1),
\end{array}\right.
\end{eqnarray}
where $\varepsilon=\pm 1$, $s(\le (n-1)/2)$ is a positive integer, and $l$ is a constant with the dimension of length.
The Schwarzschild-Tangherlini-type black hole in this theory was investigated in~\cite{DCBH,DCBH2}.
Using the binomial theorem:
\begin{align}
(x+y)^s=\sum_{p=0}^{s}
\left(
\begin{array}{cc}
s \\
p
\end{array}
\right)x^py^{s-p},
\end{align}
we obtain $m_{\rm L}$ in a very simple form as 
\begin{align}
m_{\rm L} = \frac{(n-2)V_{n-2}^{(k)}l^{2s-2}r^{n-1-2s}}{2\varepsilon\kappa_n^2}\biggl(k-(Dr)^2+\frac{r^2}{l^2}\biggl)^s.
\end{align}  
Putting the metric (\ref{vacuum}) and $m_{\rm L}=0$, we obtain the effective cosmological constant ${\tilde\lambda}$ for the $s$th-order degenerated vacuum as
\begin{align}
{\tilde\lambda}=-\frac{1}{l^2},\label{lamnbda-dc}
\end{align}  
which implies that the unique maximally symmetric vacuum is AdS (dS) for real (pure imaginary) $l$.
Hence, the following positive (or negative) mass theorem is shown independent of $k$ in this case.
\begin{Prop}
\label{th:DCmass}
({\it Signature of mass in Lovelock gravity with a unique (A)dS vacuum.}) 
In Lovelock gravity with a unique AdS (dS) vacuum and even $s$, $\varepsilon m_{\rm L}$ is non-negative (non-positive) in the domain of $r\ge 0$.
\end{Prop}
%


The Schwarzschild-Tangherlini-type vacuum solution~\cite{DCBH,DCBH2} is given by
\begin{align}
f(r) = k-\biggl(\frac{2\varepsilon\kappa_n^2M}{(n-2)V_{n-2}^{(k)}l^{2s-2}r^{n-1-2s}}\biggl)^{1/s}+\frac{r^2}{l^2}.
\end{align}  
where $m_{\rm L}=M$ is constant.
For even $s$, $\varepsilon M \ge (\le)0$ is required for real (pure imaginary) $l$ in order for $f(r)$ to be real.

The unique AdS or dS vacuum is maximally degenerated for $n=(n-1)/2$ and $n=(n-2)/2$ in odd and even dimensions, respectively.
This theory is called the dimensionally continued gravity~\cite{DCBH}, and particularly Chern-Simons gravity in odd dimensions~\cite{zanelli2005}.
The following Corollary is given from Proposition~\ref{th:DCmass}.
\begin{Coro}
\label{coro:dc}
In Lovelock gravity with a maximally degenerated AdS (dS) vacuum, $\varepsilon m_{\rm L}$ is non-negative (non-positive) in the domain of $r\ge 0$ for $n=4q+1$ or $n=4q+2$, where $q(\ge 1)$ is an integer.
\end{Coro}


\section{Schwarzschild-Tangherlini-type solution with electric charge}
\label{chargedBH}
We consider the Maxwell field as a matter field, of which action and the energy momentum are respectively given by
\begin{align}
I_{\rm matter}=&-\frac{1}{\zeta^2}\int d^nx\sqrt{-g}F_{\mu\nu}F^{\mu\nu},\\
T_{\mu\nu}=&\frac{1}{\zeta^2}\biggl(F_{\mu\rho}F_{\nu}^{~\rho}-\frac14 g_{\mu\nu}F_{\rho\sigma}F^{\rho\sigma}\biggl), \label{n-em}
\end{align}
where $\zeta$ is a real coupling constant.
For the Schwarzschild-Tangherlini-type metric (\ref{f-vacuum}), the electric solution of the Maxwell equation $\nabla_\nu F^{\mu\nu}=0$ is given by
\begin{align}
A_{\mu}dx^\mu=-\frac{Q}{r^{n-3}}dt,
\end{align}
which implies that the Faraday tensor reads
\begin{align}
F_{\mu \nu }d x^\mu \wedge d x^\nu =-\frac{(n-3)Q}{r^{n-2}}dt \wedge dr,
\end{align}
where $Q$ is a real constant.
The energy-momentum tensor is then given by
\begin{align}
T^a_{~~b}=-\frac{(n-3)^2Q^2}{2\zeta^2r^{2(n-2)}}\delta^a_{~~b}, \qquad T^i_{~~j}=\frac{(n-3)^2Q^2}{2\zeta^2r^{2(n-2)}}\delta^i_{~~j}.
\end{align}

The $r$ component of Eq.~(\ref{ufl-D}) is integrated to give the master equation for the metric function $f(r)$ as
\begin{align}
\frac{(n-2)V_{n-2}^{(k)}}{2\kappa_n^2}\sum_{p=0}^{[n/2]}{\tilde \alpha}_{(p)}r^{n-1-2p}[k-f(r)]^p=&-\frac{(n-3)V_{n-2}^{(k)}Q^2}{2\zeta^2r^{n-3}}+M,\label{f-e}
\end{align}  
where $M$ is the mass parameter~\cite{zegers2005}.
This equation means that by the following replacement:
\begin{align}
M \to M-\frac{(n-3)V_{n-2}^{(k)}Q^2}{2\zeta^2r^{n-3}},
\end{align}  
we easily obtain the electrically charged solution from the Schwarzschild-Tangherlini-type vacuum solution in Lovelock gravity.

It is shown that the asymptotic behavior of the metric function around $r=0$ is given from Eq.~(\ref{f-e}) as
\begin{align}
f(r)\simeq -\biggl(-\frac{(n-3)\kappa_n^2Q^2}{(n-2)\zeta^2{\tilde \alpha}_{(s)}}\biggl)^{1/s}\frac{1}{r^{2(n-2-s)/s}},
\end{align}  
where $s$ is the largest integer with non-zero ${\tilde \alpha}_{(s)}$ in $1 \le s \le [n/2]$.
Hence, we show the following.
\begin{Prop}
\label{th:r=0-e}
({\it Unphysical central region with electric charge.}) 
Let $s$ be the largest integer with non-zero ${\tilde \alpha}_{(s)}$ in $1 \le s \le [n/2]$.
In the Schwarzschild-Tangherlini-type spacetime with electric charge, if $s$ is even and ${\tilde \alpha}_{(s)}>0$, the metric is complex around $r=0$.
\end{Prop}

\bigskip

The above proposition confirms the fact that the metric becomes imaginary around $r$ independent of the mass parameter in Einstein-Gauss-Bonnet gravity ($s=2$) with electric charge~\cite{tm2005b}.


\begin{thebibliography}{99}
\bibitem{er2002}
R.~Emparan and H.S.~Reall,
Phys.\ Rev.\ Lett.\  {\bf 88}, 101101 (2002).
\bibitem{lovelock}
D. Lovelock,
J. Math. Phys. {\bf 12}, 498 (1971).
\bibitem{bentobertolami1996}
M.C.~Bento and O.~Bertolami,
Phys. Lett. {\bf B368}, 198 (1996).
\bibitem{Gross} 
   D. J.~Gross and E.~Witten, 
   Nucl. Phys. {\bf B277}, 1 (1986);
   D. J.~Gross and J. H.~Sloan, 
   Nucl. Phys. {\bf B291}, 41 (1987);
   R. R.~Metsaev and A. A.~Tseytlin, Phys. Lett. B {\bf 191}, 354 (1987);
  B.~Zwiebach,
   Phys. Lett. B {\bf 156}, 315 (1985);
   R. R.~Metsaev and A. A.~Tseytlin, 
   Nucl. Phys. {\bf B293}, 385 (1987).
\bibitem{zanelli2005}
J.~Zanelli, {\it Lecture notes on Chern-Simons (super-)gravities. Second edition
(February, 2008)}, arXiv:hep-th/0502193.
\bibitem{zegers2005} 
R.~Zegers, 
J. Math. Phys. {\bf 46}, 072502 (2005). 
\bibitem{whitt1988} 
B.~Whitt, 
Phys. Rev. D {\bf 38}, 3000 (1988). 
\bibitem{lovelock-thermo}
R.C.~Myers and J.Z.~Simon, Phys. Rev. D {\bf 38}, 2434 (1988);
T.~Jacobson and R.C.~Myers,
Phys. Rev. Lett. {\bf 70}, 3684 (1993)
M.~Ba\~nados, C.~Teitelboim, and J.~Zanelli, Phys. Rev. Lett. {\bf
72}, 957 (1994);
J.P.~Muniain and D.D.~Piriz,
Phys. Rev. D {\bf 53}, 816 (1996);
E.~Abdalla and L.A.~Correa-Borbonet,
Phys. Rev. D {\bf 65}, 124011 (2002).
\bibitem{cai2004}
R.-G.~Cai, Phys. Lett. {\bf B582}, 237 (2004).
\bibitem{lovelockreview} 
  C.~Garraffo and G.~Giribet,
  Mod.\ Phys.\ Lett.\  A {\bf 23}, 1801 (2008);
C.~Charmousis,
  Lect.\ Notes Phys.\  {\bf 769}, 299 (2009).
\bibitem{ck2005}
R.-G.~Cai and S.P.~Kim,
JHEP {\bf 0502}, 050 (2005).
\bibitem{ac2007}
M.~Akbar and R.-G.~Cai,
Phys. Rev. D {\bf 75},  084003 (2007).
\bibitem{cc2007}
R.-G.~Cai and L.-M.~Cao,
Phys. Rev. D {\bf 75}, 064008 (2007).
\bibitem{cchk2008}
R.-G.~Cai, L.-M.~Cao, Y.-P.~Hu, and S.P.~Kim,
Phys. Rev. D {\bf 78}, 124012 (2008).
\bibitem{Padmanabhan:2002jr}
  T.~Padmanabhan and A.~Paranjape,
  Phys.\ Rev.\  D {\bf 75}, 064004 (2007);
\bibitem{paddy2009}
T.~Padmanabhan, 
Rept. Prog. Phys. {\bf 73}, 046901 (2010).
\bibitem{ts2009}
T.~Takahashi and J.~Soda,
Phys. Rev. D {\bf 79}, 104025 (2009);
T.~Takahashi and J.~Soda,
Phys. Rev. D {\bf 80}, 104021 (2009);
T.~Takahashi and J.~Soda, 
Prog. Theor. Phys. {\bf 124}, 711 (2010);
T.~Takahashi and J.~Soda,
Prog. Theor. Phys. {\bf 124}, 911 (2010).
\bibitem{takahashi2011}
T.~Takahashi,
e-Print: arXiv:1102.1785 [gr-qc]. 
\bibitem{mp} 
R.C.~Myers and M.J.~Perry, 
Annals Phys. {\bf 172}, 304, (1986). 
\bibitem{rotateCS} 
A.~Anabalon, N.~Deruelle, Y.~Morisawa, J.~Oliva, M.~Sasaki, D.~Tempo, and R.~Troncoso, 
Class. Quant. Grav. {\bf 26}, 065002 (2009).
\bibitem{maeda2006b} 
H.~Maeda, 
Phys. Rev. D {\bf 73}, 104004 (2006). 
\bibitem{mn2008}
H.~Maeda and M.~Nozawa,
Phys. Rev. D{\bf 77}, 064031 (2008).
\bibitem{nm2008}
M.~Nozawa and H.~Maeda,
Class. Quant. Grav. {\bf 25}, 055009 (2008).
\bibitem{maeda2008}
H.~Maeda, 
Phys. Rev. D {\bf 78}, 041503(R) (2008).
\bibitem{maeda2010}
H.~Maeda, 
Phys. Rev. D {\bf 81}, 124007 (2010).
\bibitem{wald}
R.M.~Wald, {\it General Relativity}, (University of Chicago Press,
1984).
\bibitem{ce2011} 
X.O.~Camanho and J.D.~Edelstein, 
e-Print: arXiv:1103.3669 [hep-th]. 
\bibitem{km2006}
D.~Kastor and R.B.~Mann,
JHEP {\bf 0604}, 048 (2006).
\bibitem{ms1964} 
C. W.~Misner and D. H.~Sharp, 
Phys. Rev. {\bf 136}, B571 (1964).
\bibitem{tw2011}
Yu~Tian and X.-N.~Wu,
JHEP {\bf 1101}, 150 (2011).
\bibitem{hayward1998}
S. A. Hayward, 
Class.\ Quant.\ Grav.\ {\bf 15}, 3147 (1998).
\bibitem{kodama1980} 
H. Kodama, 
Prog. Theor. Phys. {\bf 63}, 1217 (1980).
\bibitem{av2010} 
G.~Abreu and M.~Visser,
Phys. Rev. D {\bf 82}, 044027 (2010).
\bibitem{kastor2008}
D.~Kastor,
Class. Quant. Grav. {\bf 25}, 175007 (2008).
\bibitem{error} 
Though the proof is correct, the statement of Proposition 8 in~\cite{nm2008} is erroneously stated.
\bibitem{bdw} 
D.~G.~Boulware, and S.~Deser,
Phys.\ Rev.\ Lett.\ \textbf{55}, 2656 (1985);
J.~T.~Wheeler,
Nucl.\ Phys.\ \textbf{B268}, 737 (1986);
D.~Lorenz-Petzold, 
Mod. Phys. Lett. {\bf A3}, 827 (1988);
R.-G.~Cai,
Phys. Rev. D {\bf 65}, 084014 (2002);
R.-G.~Cai and Qi~Guo,
Phys. Rev. D {\bf 69}, 104025 (2004).
\bibitem{tm2005}
T.~Torii and H.~Maeda,
Phys. Rev. D{\bf 71}, 124002 (2005).
\bibitem{ds2005} 
M.H.~Dehghani and M.~Shamirzaie, 
Phys. Rev. D {\bf 72}, 124015 (2005);
M.H.~Dehghani and R.B.~Mann,
Phys. Rev. D {\bf 73}, 104003 (2006).
\bibitem{DCBH}
  M.~Ba\~nados, C.~Teitelboim and J.~Zanelli,
  Phys.\ Rev.\  D {\bf 49}, 975 (1994);
R.-G.~Cai and K.-S.~Soh,
Phys. Rev. D {\bf 59}, 044013 (1999).
\bibitem{DCBH2}
J.~Crisostomo, R.~Troncoso, and J.~Zanelli,
Phys. Rev. D {\bf 62}, 084013 (2000);
R.~Aros, R.~Troncoso, and J.~Zanelli,
Phys. Rev. D {\bf 63}, 084015 (2001).
\bibitem{hayward1994} 
S.~A.~Hayward, 
Phys. Rev. D {\bf 49}, 6467 (1994).
\bibitem{hayward1996}
S. A. Hayward, 
Phys. Rev. D. {\bf 53}, 1938 (1996).
\bibitem{typo} 
The factor $2$ in the left-hand side in (3.10) in~\cite{mn2008} is a typo.
\bibitem{df1990}
N.~Deruelle and L.~Farina-Busto,
Phys. Rev. D {\bf 41}, 3696 (1990). 
\bibitem{osj2011}
S.~Ohashi, T.~Shiromizu, and S.~Jhingan,
e-Print: arXiv:1103.3826 [gr-qc].
\bibitem{vaidya-L}
M.~Nozawa and H.~Maeda,
Class. Quant. Grav. {\bf 23}, 1779 (2006).
\bibitem{Garraffo:2007fi}
  C.~Garraffo, G.~Giribet, E.~Gravanis and S.~Willison,
  J.\ Math.\ Phys.\  {\bf 49}, 042502 (2008).
\bibitem{Gravanis:2010zs}
E.~Gravanis, 
Phys. Rev. D {\bf 82}, 104024 (2010).
\bibitem{Oliva:2011xu}
  J.~Oliva and S.~Ray,
  arXiv:1104.1205 [gr-qc].
\bibitem{wheeler-lovelock} 
J.~T.~Wheeler,
Nucl.\ Phys.\ \textbf{B273}, 732 (1986).
\bibitem{cd2002} 
C.~Charmousis and J-F.~Dufaux, 
Class.\ Quant.\ Grav.\ {\bf 19}, 4671 (2002).
\bibitem{wormhole-L}
G.~Dotti, J.~Oliva, and R.~Troncoso,
Phys. Rev. D {\bf 75}, 024002 (2007);
G.~Dotti, J.~Oliva, and R.~Troncoso,
Phys. Rev. D {\bf 76}, 064038 (2007);
F.~Canfora and A.~Giacomini,
Phys. Rev. D {\bf 78}, 084034 (2008).
\bibitem{dot2010} 
G.~Dotti, J.~Oliva, and R.~Troncoso, 
Phys. Rev. D {\bf 82}, 024002 (2010).
\bibitem{Lifshitz-L}
M.H.~Dehghani and R.B.~Mann,
JHEP {\bf 1007}, 019 (2010).
\bibitem{wald1993}
R.M.~Wald, 
Phys. Rev. D {\bf 48}, R3427 (1993).
\bibitem{iyerwald1994}
V. Iyer and R. M. Wald,
Phys. Rev. D {\bf 50}, 846 (1994).
\bibitem{jkm1994}
T.~Jacobson, G.~Kang, and R.C.~Myers, 
Phys. Rev. D {\bf 49}, 6587 (1994).
\bibitem{Hawking:1982dh}
  S.~W.~Hawking and D.~N.~Page,
  Commun.\ Math.\ Phys.\  {\bf 87}, 577 (1983).
\bibitem{Kastor}
  D.~Kastor, S.~Ray and J.~Traschen,
 ``Mass and Free Energy of Lovelock Black Holes'' (to appear).
\bibitem{Kastor:2009wy}
  D.~Kastor, S.~Ray and J.~Traschen,
  Class.\ Quant.\ Grav.\  {\bf 26}, 195011 (2009).
\bibitem{Kastor:2010gq}
  D.~Kastor, S.~Ray and J.~Traschen,
  Class.\ Quant.\ Grav.\  {\bf 27}, 235014 (2010).
\bibitem{dg2005} 
G.~Dotti and R.J.~Gleiser, 
   Phys. Lett. {\bf B627}, 174 (2005).
\bibitem{tm2005b}
T.~Torii and H.~Maeda,
Phys. Rev. D{\bf 72}, 064007 (2005).


\end{thebibliography}
\end{document}